\documentclass[acmsmall]{acmart}
\usepackage{graphicx}
\usepackage{array}
\usepackage{tikz}
\usepackage{lineno}
\usepackage{amsmath}
\usepackage{mathrsfs}
\usepackage{tabularx}
\usepackage{lscape}
\usepackage{booktabs}
\usepackage{multirow}
\usepackage{mathtools}
\usepackage{float}
\usepackage{geometry}
\AtBeginDocument{%
  \providecommand\BibTeX{{%
    \normalfont B\kern-0.5em{\scshape i\kern-0.25em b}\kern-0.8em\TeX}}}

\setcopyright{acmcopyright}
\copyrightyear{2020}
\acmYear{2020}
\acmDOI{10.1145/1122445.1122456}

\acmJournal{JACM}
\acmVolume{37}
\acmNumber{4}
\acmArticle{111}
\acmMonth{8}

\begin{document}

\title{Security and Privacy for Artificial Intelligence: Opportunities and Challenges}


\author{Ayodeji Oseni}
\email{ayodeji.s.oseni@gmail.com}
\authornotemark[1]
\affiliation{%
  \institution{The University of New South Wales @ ADFA}
  \city{Canberra}
  \country{Australia}
}
\author{Nour Moustafa}
\authornotemark[1]
\email{nour.moustafa@unsw.edu.au}
\affiliation{%
  \institution{The University of New South Wales @ ADFA}
  \city{Canberra}
  \country{Australia}
 }

\author{Helge Janicke}
 \email{helge.janicke@cybersecuritycrc.org.au}
\affiliation{%
  \institution{The Cyber Security Cooperative Research Centre (CSCRC)}
  \city{Perth}
  \state{Beijing Shi}
  \country{Australia}
  }

\author{Peng Liu}
\email{pliu@ist.psu.edu}
\affiliation{%
  \institution{Penn State, University Park}
  \city{Pennsylvania}
  \country{USA}
}
\author{Zahir Tari}
\email{zahir.tari@rmit.edu.au}
\affiliation{%
  \institution{RMIT University}
  \city{Melbourne}
  \country{Australia}
}

\author{Athanasios Vasilakos}
\email{Athanasios.Vasilakos@uts.edu.au}
\affiliation{%
  \institution{The University of Technology Sydney}
  \city{Sydney}
  \country{Australia}
}

\renewcommand{\shortauthors}{Oseni and Moustafa, et al.}

\begin{abstract}
The increased adoption of Artificial Intelligence (AI) presents an opportunity to solve many socio-economic and environmental challenges; however, this cannot happen without securing AI-enabled technologies. In recent years, most AI models are vulnerable to advanced and sophisticated hacking techniques. This challenge has motivated concerted research efforts into adversarial AI, with the aim of developing robust machine and deep learning models that are resilient to different types of adversarial scenarios. In this paper, we present a holistic cyber security review that demonstrates adversarial attacks against AI applications, including aspects such as adversarial knowledge and capabilities, as well as existing methods for generating adversarial examples and existing cyber defence models. We explain mathematical AI models, especially new variants of reinforcement  and federated learning, to demonstrate how  attack vectors would exploit vulnerabilities of AI models. We also propose a systematic framework for demonstrating attack techniques against AI applications, and reviewed several cyber defences that would protect the AI applications against those attacks. We also highlight the importance of understanding the adversarial goals and their capabilities, especially the recent attacks against industry applications, to develop adaptive defences that assess to secure AI applications. Finally, we describe the main challenges and future research directions in the domain of security and privacy of AI technologies. 
\end{abstract}



\keywords{Adversarial Attacks, Adversarial Examples, Deep Learning, Machine Learning, Privacy, Security}

\maketitle

\section{Introduction}
\label{S:1}
Recent advances in technology, coupled with growth in computational capacities, have led to the adoption of Artificial Intelligence (AI) techniques in several applications \cite{papernot2018security}. For example, machine learning models are being used to drive innovation in the area of health care, gaming and finance, while autonomous car manufacturers rely on the deep learning models to create pipelines for self-driving cars. Machine learning (ML) models and, more recently, deep learning (DL) algorithms used in several AI systems today make it possible to automate tasks and processes, thereby introducing new capabilities and functionalities that were not previously possible. For instance, in 2019, DeepMind's AlphaStar, an AI system based on deep reinforcement learning, reached the Grandmaster level in the video game, StarCraft II, by beating several high-level human players \cite{risi2020behind}. 

Despite the noticeable success and benefits of using machine learning, many of the machine learning models in use today are vulnerable to several adversarial examples, where adversaries seek to violate the confidentiality, integrity or availability of machine learning models by using inputs that are specifically crafted to cause the models to make false predictions \cite{barreno2006, munoz2019challenges}. In many cases, AI systems are designed with no considerations for security, making them highly vulnerable to adversarial examples. Adversarial attacks on AI systems can occur either at the training or testing phase of machine learning \cite{liu2018survey}. At the training phase, an adversary can inject malicious data into a training dataset to manipulate input features or data labels. These attacks, referred to as poisoning attacks in literature, can easily be carried out in applications that use training data from untrusted sources. \citet{barreno2006} demonstrated how an adversary, with knowledge of a machine learning model, can change the original distribution of a training dataset by modifying the training data. Attacks at the testing phase, known as evasion attacks, are the most prevalent type of attacks on machine learning models as they exploit the vulnerabilities of a model to generate adversarial examples, which are then used to evade the model at test time \cite{liu2018survey}.

The adoption of AI techniques in many applications presents a unique opportunity to solve many socio-economic and environmental challenges. However, this cannot happen without focused research on securing these technologies. The field of adversarial machine learning has recently been receiving significant research interest, and continuing focus in this area will undoubtedly ensure the ubiquitous spread of transformative AI technologies. As AI becomes increasingly integrated into different aspects of human activities and lifestyles, robust algorithms are essential to the progression of a protected and safe innovative future.

\textbf{Motivation} -- Adversarial examples make the machine learning models used in many AI systems vulnerable to adversarial attacks. An adversarial attack on the confidentiality of a machine learning model would seek to expose the model structure, the training or test dataset \cite{papernot2018security}, thereby impacting the privacy of data sources. Adversarial examples could also be used to attack the input integrity of machine learning models, thereby exploiting the imperfections made by the learning algorithm during the model's training phase \cite{papernot2016transferability}. Attacks on availability fall within the realms of adversarial behaviours which prevents legitimate users from being able to access a model's outputs or features. The security and privacy of AI systems against cyber adversaries have individually attracted attention in the research community over the last few years \cite{chen2019lightweight,papernot2018security,pitropakis2019taxonomy,ren2020adversarial,zhang2020adversarial}. However, there has been minimal effort to review the literature and experimental results that illustrate a comparative analysis of AI security and privacy, as we present in this paper.

\textbf{Our Contributions} -- We summarize the main contributions of our work in five aspects:
\begin{itemize}
    \item We present a detailed analysis of previous related surveys. This constitutes a contribution to literature through the identification of limitations in previous surveys related to the development of secure AI applications, as we addressed in this work.
    \item We present a brief overview of machine learning task categories, including deep learning and federation learning
    \item We present a new adversarial attack framework that illustrates advanced attacks that would exploit AI applications and measures their threats.
   \item We present a new defence framework that demonstrates cyber defence methods for protecting AI systems against adversarial attacks
    \item We discuss challenges in this domain and provide suggestions on future research directions. 
\end{itemize}


The remainder of this paper is structured as follows: First, in Section \ref{reviews}, we explain the most recent surveys and reviews that attempt to explain security or privacy insights of AI applications. Second, we provide an overview of machine and deep learning task categories while also considering federated learning in Section \ref{machine-learning}. Third, we discuss the most complete set of attack types that would breach AI applications, including real-world attack scenarios and motives, in Section \ref{attacks-ai}. Fourth, we explain in Section \ref{defenses-ai} defense methods and techniques that would protect AI models against cyber adversaries. 
Lastly, we explain lessons learned and future research directions based on our systematic review analysis, in section \ref{challenges-future}.

\section{Most Recent Reviews}
\label{reviews}

To develop an overview of academic activity relating to security and privacy of AI technologies, we identify and review the most recent surveys and reviews conducted between 2018 and 2020 as presented in Table \ref{table:1}. A total of 16 high-quality systematic reviews were identified; 11 of which were published in peer-reviewed journals such as IEEE Access, Elsevier, ACM, Applied Sciences and ScienceDirect. An analysis of these reviews is presented as follows.
 
 From a data-driven view, \citet{liu2018survey} presented a systematic survey on the existing security threats and corresponding defensive techniques during the training and testing phases of machine learning. Given the lack of a comprehensive literature review covering the security threats and defensive techniques during the two phases of machine learning, their foundational work provided a detailed summary of existing adversarial attack techniques against machine learning and countermeasures. Particularly, they presented a detailed description of machine learning and introduced the concept of adversarial machine learning. While their survey has informed further research into adversarial machine learning, the authors did not review existing security threats on reinforcement learning.

\begin{table*}
\caption{Recent Surveys and Reviews in the area of Security and Privacy in AI} \;
\centering
\tiny
\setlength\tabcolsep{4pt}
\begin{tabularx}{\linewidth}{|>{\hsize=0.5\hsize}X|
                              >{\centering \hsize=0.4\hsize}X|
                              >{\centering \hsize=0.4\hsize}X|
                              >{\hsize=0.4\hsize}X|
                              >{\hsize=1.0\hsize}X|
                              >{\hsize=0.9\hsize}X|}
\hline
    \textbf{Reference} & \textbf{Publication Date} & \textbf{Main Area of Focus} & \textbf{Research Method} & \textbf{Main Contributions of Work} & \textbf{Limitations of Work} \\
\hline
\citet{liu2018survey} & February 2018 & Machine Learning & Systematic review & Reviewed a cross-section of security threats and defensive techniques during stages of ML from a \textit{data-driven} view & Review did not include existing threats on Reinforcement Learning \\

\hline
\citet{akhtar2018threat} & February 2018 & Deep Learning & Systematic review & Provided a comprehensive survey of adversarial attacks on DL in computer vision & Restrictive context of computer vision \\

\hline
\citet{papernot2018security} & April 2018 & Supervised Learning (Classification) & Systematic review & Cataloging of attacks and defenses in ML. Introduced a unified threat model in describing security and privacy issues in ML systems & Article primarily focused on attacks and defenses relating to ML classification \\ 

\hline
\citet{thomas2018adversarial} & July 2018 & Machine Learning & Semi-systematic review & Surveyed the current landscape of research in the area of adversarial machine learning with analysis of results and the trends in research & Lacks the depth of most systematic review papers in this domain \\ 

\hline
\citet{bae2018security} & July 2018 (\textit{pre-print}) & Deep Learning & Systematic review & Surveyed possible attacks and current defense methods on DL. Taxonomized approaches to Privacy in DL & Survey paper only focused on security and privacy issues in DL \\

\hline
\citet{chakraborty2018adversarial} & September 2018 (\textit{pre-print}) & Deep Learning & Systematic review & Provided a summary of recent advances in adversarial attacks with their countermeasures. & Paper did not address ways in which countermeasures can be improved \\ 

\hline
\citet{li2018security} & October 2018 (\textit{pre-print}) & Deep Learning & Systematic review & Reviewed some of the attack and defensive strategies in deep neural networks & Selective review of a small fraction of attack and defensive strategies in deep neural networks and mainly in context of computer vision. \\ 

\hline
\citet{ozdag2018adversarial} & November 2018 & Deep Learning & Systematic review & Reviewed types of adversarial attacks and defenses in deep neural networks & Survey paper lacks the depth of some of the other reviews on adversarial attacks in the DL domain \\ 

\hline
\citet{biggio2018wild} & December 2018 & Machine Learning & Systematic review & Reviewed works focusing on security of machine learning, pattern recognition, and deep neural networks & Review only focused on adversarial attacks in the context of computer vision and cybersecurity \\

\hline
\citet{qiu2019review} & March 2019 & Deep Learning & Systematic review & Reviewed recent studies on AI adversarial attack and defense technologies & Work did not cover some of the privacy attacks on traditional machine learning algorithms such as Naive Bayes, decision trees and random forests \\

\hline
\citet{wang2019security} & August 2019 & Machine learning & Systematic Review & Review of the security properties of machine learning algorithms in adversarial settings & Work only reviewed a small fraction of adversarial attacks against ML systems  \\

\hline
\citet{wiyatno2019adversarial} & November 2019 (\textit{pre-print}) & Supervised Learning (Classification) & Systematic review & In-dept review of adversarial attacks, defenses and discussion of the strengths and weaknesses of each method & Paper only focused on adversarial attack methods and defenses which apply to ML classification tasks \\ 

\hline
\citet{pitropakis2019taxonomy} & November 2019 & Machine Learning & Systematic review & Provided a comprehensive taxonomy and systematic review of attacks against machine learning systems & Paper did not include review of defenses against the adversarial attacks \\

\hline
\citet{li2020adversarial} & February 2020 & Supervised Learning & Systematic review & Provides a complimentary summary of adversarial attacks and defenses for Cyber-Physical systems beyond the field of computer vision & Survey paper was based solely on cyber-physical systems \\ 

\hline
\citet{ren2020adversarial} & March 2020 & Deep Learning & Systematic review & Provided summarized review of adversarial attacks and defensive techniques in deep learning & Paper focused mainly on attacks and defenses in the context of computer vision \\

\hline
\citet{zhang2020adversarial} & April 2020 & Deep Learning & Systematic review & Reviewed research efforts on generating adversarial examples on textual deep neural networks & Survey paper did not cover many of the defensive techniques for attacks on deep learning models \\ 

\hline
\end{tabularx}
\label{table:1}
\end{table*}

The first comprehensive survey in the domain of adversarial attacks on deep learning in computer vision was undertaken by \citet{akhtar2018threat}. Reviewing existing and proposed defenses for adversarial attack methods, this survey includes a discussion of adversarial attack methods that have been used successfully against deep neural networks in both ’laboratory settings’ and real-world scenarios. This work is however limited as it presents the most ’influential’ and ’interesting’ attacks on deep learning in the restricted context of computer vision.

Based on a growing recognition that machine learning models are increasingly vulnerable to a wide range of adversarial capabilities, \citet{papernot2018security} carried out a systematic review of ML security and privacy with specific focus on adversarial attacks on machine learning systems and their countermeasures. In this article, the authors considered ML threat model from the perspective of a data pipeline and reviewed recent attacks on ML and their defenses at the training and testing stages. Their review also covered existing works in the area of differential privacy. A review of the current landscape of research in the area of adversarial machine learning  presented by \citet{thomas2018adversarial} analyses the results and trends in adversarial machine learning research. However, this work did not specifically focus on any of the attacks on ML systems and it lacks the depth of many review papers in this domain.

On the security and privacy of deep learning systems in several applications, \citet{bae2018security} presents a systematic review of recent research. Covering the foundations of deep learning and some of the privacy-preserving techniques in literature, the article introduces the concept of Secure AI and provides an extensive analysis of many of the attacks that have successfully been used against deep learning as well as their defense techniques. Approaches to privacy in deep learning are taxonomized and future research directions are presented based on identified gaps. A review of different types of adversarial attacks and their defences reported by \citet{chakraborty2018adversarial} provides a summary of adversarial attack types and proposes defenses through analysis of different threat models and attack scenarios. The attacks and countermeasures reviewed were not restrictive to specific deep learning applications. Particularly, the need for robust deep learning architectures is suggested as countermeasures against adversarial attacks. However, the article did not provide steps in which this can be achieved. Such steps might have been included in future research directions.

An introduction into the foundations of adversarial machine learning presented by \citet{li2018security} focuses on recent attack and defensive strategies in deep neural networks. This work describes some of the metrics used in literature for measuring perturbations of input samples. Whilst including adversarial attacks in reinforcement learning, the work only covered a small fraction of the attack and defensive strategies in deep neural networks that have been proposed in literature. A review of some of the well-known adversarial attacks and defenses against deep neural networks provided by \citet{ozdag2018adversarial} reports some of the solution models and results presented in the NIPS 2017 Adversarial Learning competition organized by Google Brain. Though showcasing relevant information, the survey did not cover many of the adversarial attacks that have been demonstrated against deep neural networks.

\citet{biggio2018wild} in their paper provided an overview of the state of research in the area of adversarial machine learning covering a period of about ten years mainly in the context of computer vision and cybersecurity tasks. The paper covers foundations of deep learning threat models, attack types and their timeline, as well as misconceptions relating to the security evaluation of machine learning algorithms. Furthermore, the paper suggests the need for research into the design of machine learning models that are against adversarial attacks. Similar to other review papers, \citet{qiu2019review} presented a comprehensive review of recent research progress on adversarial attacks against deep learning and their defenses. The review focused on attack methods at both the training and testing stages of machine learning in the contexts of computer vision, image classification, natural language processing, cybersecurity and the cyber-physical world. The review also includes a number of proposed defensive strategies in literature. While this work comprehensively reviewed several demonstrated attacks against AI systems and their defenses, it did not cover some of the privacy issues associated with traditional machine learning algorithms.

A review of the security properties of machine learning algorithms in adversarial settings presented by \citet{wang2019security} is also similar to other recent reviews in this area, covering foundations of adversarial machine learning. Some of the adversarial attacks that have been demonstrated in literature as well as their countermeasures are also included. A comprehensive review of adversarial attack methods against machine learning models in the visual domain is reported by \citet{wiyatno2019adversarial} with specific focus on the application of adversarial examples to machine learning classification tasks. Discussing the strengths and weaknesses of each of the reviewed adversarial attacks and defences, the review covered many of the attacks against machine learning systems in literature. Our work significantly expands their work with an analysis of attacks on unsupervised and reinforcement learning, which is not included in \cite{wiyatno2019adversarial}.

A taxonomy and review of attacks against machine learning systems undertaken by \citet{pitropakis2019taxonomy} categorizes attacks based on their key characteristics in order to understand the existing attack landscape towards proposing appropriate defenses. This work adequately covers many of the adversarial attacks on machine learning in the context of intrusion detection, spam filtering, visual recognition and other applications. A limitation of this work however is that it did not include review of defenses against the adversarial attacks. A review of adversarial attacks and potential defenses in non-camera sensor-based cyber-physical systems by \citet{li2020adversarial} focuses on adversarial attacks on surveillance sensor data, audio data and textual input, and their defensive strategies. The work introduces a general workflow which could be used to describe adversarial attacks on cyber-physical systems, thus covering recent adversarial attacks against cyber-physical systems. It is however important to note that only classification and regression tasks are covered.

A paper on the review of adversarial attacks and defenses in deep learning by \citet{ren2020adversarial} introduces foundational concepts of adversarial attacks and then provides summarized review of adversarial attacks and defensive techniques in deep learning. While some of the adversarial attacks reviewed in this work are based on recent publications, the paper emphasises only attacks applicable to the context of computer vision. Recent research efforts focusing on adversarial attacks on deep learning models in the context of natural language processing (NLP) surveyed by \citet{zhang2020adversarial} covers the foundations of adversarial attacks and deep learning techniques in natural language processing. The survey paper includes summaries of black-box and white-box attacks methods that have been demonstrated on deep learning models in NLP. They also reviewed two common defensive techniques that have been proposed in literature for achieving robust textual deep neural networks, namely adversarial training and knowledge distillation. While this work covers most of the attacks that have been demonstrated against textual deep neural networks to date, it did not cover many of the defensive techniques that have been proposed in this area.

\section{Machine Learning Overview}
\label{machine-learning}

The idea of machine learning (ML) is not new; it has been since the 1970s when the first set of algorithms were introduced \cite{7548905}. Machine learning deals with the problem of extracting features from data in order to solve many predictive tasks; examples of which are forecasting, anomaly detection, spam filtering and credit risk assessment. Its primary goal is to predict results based on some input data. Data is a fundamental component of every machine learning system. For instance, in order to predict if an email is a spam or not, a machine would need to have been trained with samples of spam messages - the more diverse the data used in training a machine, the better the result of the prediction. Input data in machine learning is typically divided into training and test data. Training data is used in developing a machine learning model; and once the accuracy of prediction of the model is satisfactory, test data is then fed into the model.


The main components of Machine Learning are: tasks, models and features \cite{flach2012machine}. Tasks refer to the problems that can be solved using machine learning. Many machine learning models are designed to solve only a small number of tasks. Models define the output of machine learning. They are simply trained by sample data in order to process additional data for making predictions. Features are an essential component of machine learning as they are characteristics of the input data which simplifies learning of patterns between the input data and output data. Algorithms are used to solve learning problems or tasks. \citet{flach2012machine} described machine learning as the art of using right features to develop the right models used to solve a given problem. 


\begin{figure*}[htbp]
\centering\includegraphics[width=0.9\linewidth]{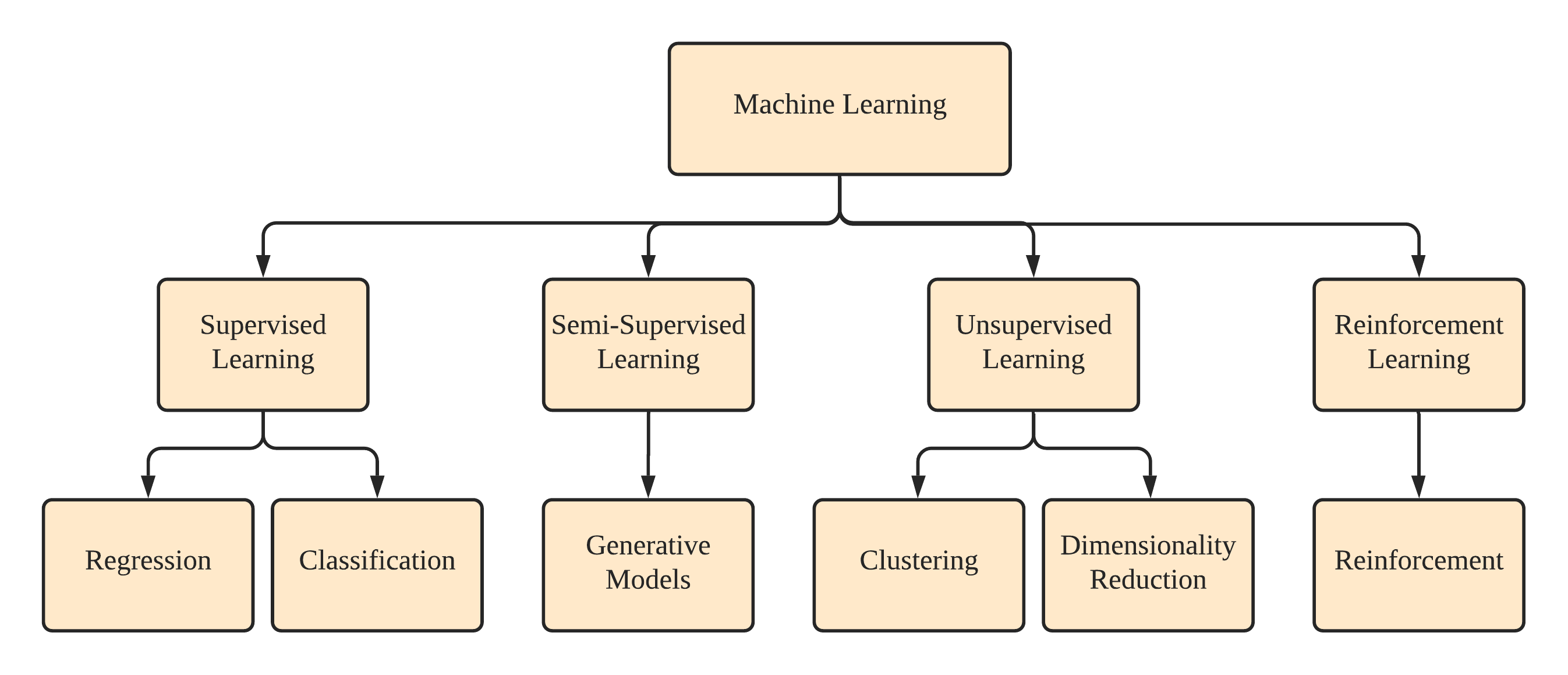}
\caption{Machine Learning Task Categories}
\label{ml_categories}
\end{figure*}

The tasks that are solved using machine learning are commonly divided into: Supervised Learning, Unsupervised Learning, Semi-supervised Learning and Reinforcement Learning \cite{lee2020machine}, as detailed below. Fig. \ref{ml_categories} shows the machine learning task categories.

\textbf{Supervised Learning} is a machine learning approach where an algorithm learns patterns from labelled training dataset and uses this for prediction or classification \cite{theobald2017machine}. In supervised learning, the training dataset is either pre-categorised or numerical. Supervised learning tasks can be grouped into either Classification and Regression techniques. On the other hand, \textbf{Unsupervised learning} deals with unlabeled input data and the learning algorithms focuses on analyzing similarities among elements of the input data to infer meaningful features \cite{theobald2017machine, lee2020machine} which are then extracted to create possible output labels. As the name implies, \textbf{Semi-supervised} learning is an approach to machine learning that combines elements of supervised learning and unsupervised learning. It extend elements of supervised learning type to include additional information typical of the unsupervised learning and vice versa \cite{zhu2009introduction}. \textbf{Reinforcement learning} is a machine learning method that enables an agent to learn in an interactive environment through trial and error using feedback that it obtains from its own actions and experiences \cite{sutton2018reinforcement, theobald2017machine}. The reinforcement learning problem involves an agent which interacts with its environment by performing certain actions and then receiving rewards for its actions. The goal of this learning approach is therefore to learn how to take actions in order to maximize the reward.

\subsection{Deep Learning Methods}
\label{S:2.5}
Deep learning is a specialized field of machine learning based on deep artificial neural networks. The machine learning methods discussed so far in this section are today referred to as Shallow Learning due to the requirement of a feature engineer to identify relevant characteristics from the input data \cite{apruzzese2018effectiveness}.  A common property of many of the shallow learning techniques is their relatively simple architecture often consisting of a single layer used to convert the input data into a problem-specific feature space \cite{deng2014tutorial}. In contrast, the more recent Deep Learning techniques rely on multi-layered representation and abstraction of the input data to perform complex learning tasks and feature selection \cite{apruzzese2018effectiveness, lee2020machine}. While many shallow learning algorithms have been effective in solving many well-constrained problems, they have been found to perform poorly in problems that involve extracting well-represented features from data such as in areas of computer vision and natural language processing \cite{deng2012three}. However, deep learning solves this problem by building multiple layers of simple features to represent a complex concept \cite{yuan2019adversarial}. The vast increase in the amount of data available nowadays and the increased chip processing abilities has played an important role in the development of deep learning architectures \cite{witten2017deep}.

\subsection{Federated Learning}
The concept of federated learning was originally introduced by Google in 2016 \cite{yang2019federated, mcmahan2016communication}. This learning approach enables the training of a centralized model on unevenly distributed data spread over federated network of nodes \cite{bonawitz2019towards, konevcny2016federated}. The primary motivation for federated learning arises from the need to train models based on data from users' mobile devices which cannot be stored centrally data centers due to privacy concerns \cite{konevcny2016optimization}. Federated learning provides distinct privacy advantages compared to the other machine learning types due to the fact that only minimal updates necessary to improve a particular model is transmitted for federated learning, and this data is dependent on the training objective. The performance of federated learning improves with more data, that is, the more the nodes available to train the model, the better the model.

\section{Attacks on AI Systems: A Holistic Review}
\label{attacks-ai}
\begin{figure*}[h]
\centering\includegraphics[width=0.7\linewidth]{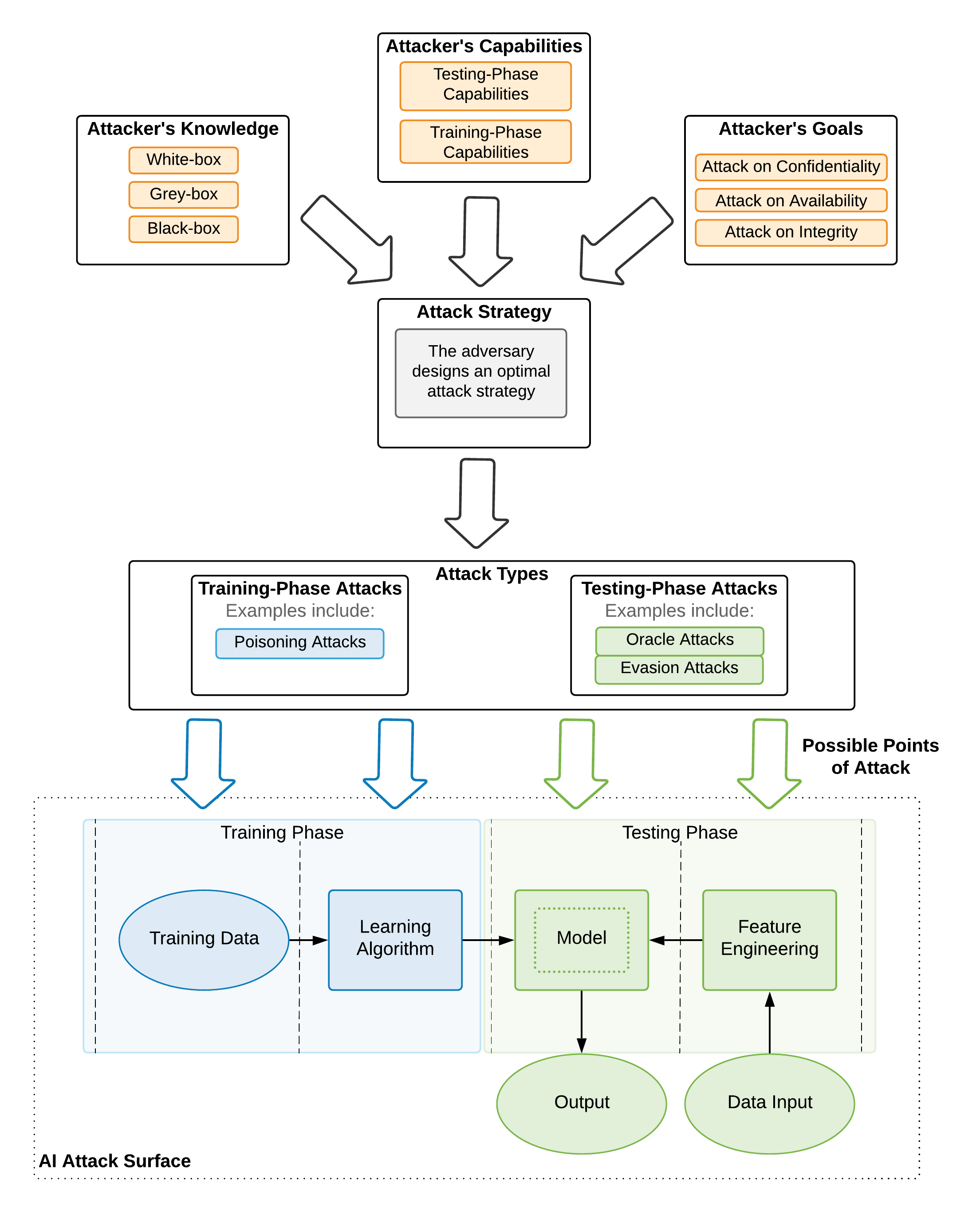}
\caption{Proposed Framework for analysis of adversarial attacks against AI models}
\label{analysis_framework}
\end{figure*}

The widespread use of AI technologies and the fact that most of these technologies are vulnerable to adversarial attacks is concerning. Adversarial attacks are often in the form of manipulations of the input to an ML or DL model with the aim of causing a misclassification of the input data. Adversarial attacks on AI models differ and depends largely on the task category to which the algorithm belongs. In other words, threats to AI systems, in general, are mostly the same, but the approaches to exploitation differ depending on the algorithm in use. In this section we first present a result of literature search covering attacks on AI systems over the last 10 years and then present a detailed analysis of the identified literature.
Literature covering attacks on AI systems were identified through Google Scholar search using key words such as Adversarial Attacks, Adversarial Examples, Adversarial Machine Learning, Attacks and Defenses on Deep Learning, Attacks and Defenses on Machine Learning, Membership Inference Attacks, Inversion Attacks, Extraction Attacks, Evasion Attacks and Poisoning Attacks. Many of the papers selected were based on works carried out between 2010 and 2020 and the results of the search are as presented in Table \ref{table:1.1} and Table \ref{table:1.2}.
To effectively assess the security and privacy of AI systems, it is important to understand the attack surface of the system, and the attack vectors that an attacker can potentially use to exploit the system. We therefore draw insights from existing frameworks \cite{barreno2010security, huang2011adversarial, munoz2017towards, biggio2014security, biggio2013security, papernot2018security} and present a new framework as a holistic approach to the quantitative analysis of adversarial attacks on AI models (Fig. \ref{analysis_framework}). 

\begin{table*}
\caption{Survey of Training Phase Attacks on AI Systems} \;
\centering
\tiny
\setlength\tabcolsep{4pt}
\begin{tabularx}{\linewidth}{|>{\hsize=0.6\hsize}X|
                              >{\centering \hsize=0.2\hsize}X|
                              >{\centering \hsize=0.3\hsize}X|
                              >{\centering \hsize=0.6\hsize}X|
                              >{\centering \hsize=0.6\hsize}X|
                              >{\centering \hsize=0.6\hsize}X|
                              >{\centering \hsize=0.6\hsize}X|
                              >{\hsize=0.6\hsize}X|}
\hline
    \textbf{Reference} & \textbf{Year} & \textbf{Attack Type} & \textbf{ML Algorithm} & \textbf{ML Application} & \textbf{Dataset} & \textbf{Attacker's Goal} & \textbf{Attacker's Knowledge}\\   
    
\hline
\citet{biggio2012poisoning} & 2012 & Poisoning & Support Vector Machine & Handwritten digit recognition & MNIST dataset &  &  \\
\hline
\citet{biggio2012adaptive} & 2012 & Poisoning & Principal Component Analysis & Adaptive biometric recognition system & 2400 face images & Integrity attack & White-box  \\
\hline
\citet{biggio2013poisoning} & 2013 & Poisoning & Principal Component Analysis & Adaptive biometric verification system & 2400 face images & Integrity attack & Grey-box  \\
 \hline
\citet{newell2014practicality} & 2014 & Poisoning & Support Vector Machine, Naive Bayes & Sentiment Analysis & Twitter 'tweets' & Integrity attack & Black-box, White-box \\
\hline
\citet{biggio2014security} & 2014 & Poisoning & Support Vector Machine & Image recognition & MNIST & Integrity attack & White-box \\
 \hline
\citet{biggio2013security} & 2014 & Poisoning & v-Support Vector Machine & Network intrusion detection & Network packets & Indiscriminate integrity attack & Grey-box \\
 \hline
\citet{mozaffari2015systematic} & 2015 & Poisoning & Best-first decision tree, Ripple-down rule learner, Naive Bayes decision tree, Nearest-neighbor classifier, Multilayer Perceptron & Healthcare & Medical dataset & Targeted integrity attack & Grey-box \\
 \hline
\citet{li2016data} & 2016 & Poisoning & Collaborative Filtering & Collaborative Filtering & MovieLens & Integrity and Availability attack & White-box \\
 \hline
\citet{munoz2017towards} & 2017 & Poisoning & Multi-Layer Perceptrons, Logistic Regression, Single-layer Artificial Neural Network, Convolutional Neural Network & Spam filtering, Malware detection, Handwritten digit recognition & Spambase, Ransomware, MNIST & Integrity attack & White-box \\
 \hline
\citet{burkard2017analysis} & 2017 & Poisoning & Support Vector Machines & Data Streams &  & Integrity attack & White-box \\
 \hline
\citet{yang2017generative} & 2017 & Poisoning & Feed-forward Neural Network, Convolutional Neural Network & Image Classification & MNIST, CIFAR-10 & Integrity attack & White-box \\
 \hline
\citet{shi2017evasion} & 2017 & Poisoning & Feed-forward Neural Network & Text and Image Classification & Reuters-21578, Flower dataset & Integrity attack & Black-box \\
 \hline
\citet{gu2017badnets} & 2017 & Backdoor & Convolutional Neural Network & Digit Recognition & MNIST & Availability attack &  \\
 \hline
\citet{jagielski2018manipulating} & 2018 & Poisoning & Linear Regression & Warfarin Dose Prediction, Loan Assessment, House pricing prediction & Healthcare dataset, Loan dataset and House pricing dataset & Poisoning availability attack & White-box, Black-box \\
 \hline
\citet{shafahi2018poison} & 2018 & Poisoning & Convolutional Neural Network & Image Classification & CIFAR-10 & Integrity attack  & Grey-box \\
 \hline
\citet{suciu2018does} & 2018 & Poisoning & Convolutional Neural Network, Linear Support Vector Machine, Random Forest & Image Recognition, Android Malware detector, Twitter-based exploit predictor, Data breach predictor & CIFAR-10, Dreblin dataset, Twitter tweets & Integrity attack & Grey-box \\
 \hline
\citet{lovisotto2019biometric} & 2019 & Poisoning & Convolutional Neural Network (FaceNet, VGG16, ResNet-50) & Biometric system & VGGFace, VGGFace2 & Integrity attack & White-box \\
 \hline
\citet{jiang2020poisoning} & 2020 & Poisoning & Convolutional Neural Network & Image Recognition & BelgiumTS, GTSRB & Integrity attack & Black-box \\
 \hline
\citet{kloft2012security} & 2020 & Poisoning & Centroid & Centroid Anomaly Detection & real HTTP traffic & Violate system integrity & White-box \\
\hline
\end{tabularx}
\label{table:1.1}
\end{table*}

\begin{table*}
\caption{Survey of Testing Phase Attacks on AI Systems} \;
\centering
\tiny
\setlength\tabcolsep{4pt}
\begin{tabularx}{\linewidth}{|>{\hsize=0.6\hsize}X|
                              >{\centering \hsize=0.2\hsize}X|
                              >{\centering \hsize=0.3\hsize}X|
                              >{\centering \hsize=1.0\hsize}X|
                              >{\centering \hsize=0.6\hsize}X|
                              >{\centering \hsize=0.5\hsize}X|
                              >{\centering \hsize=0.5\hsize}X|
                              >{\hsize=0.4\hsize}X|}
\hline
    \textbf{Reference} & \textbf{Year} & \textbf{Attack Type} & \textbf{ML Algorithm} & \textbf{ML Application} & \textbf{Dataset} & \textbf{Attacker's Goal} & \textbf{Attacker's Knowledge}\\   
    
 \hline
 \citet{szegedy2013intriguing} & 2013 & Evasion & Convolutional Neural Network (AlexNet, QuocNet), Fully Connected Network, Autoencoder & Handwritten Digit Recognition, Image Classification & MNIST, ImageNet & Misclassification & Black-box \\
 \hline
\citet{biggio2013security} & 2014 & Oracle & Support Vector Machine, Logistic Regression & Spam Filtering & TREC 2007 & Indiscriminate integrity violation & White-box \\
 \hline
 \citet{biggio2013security} & 2014 & Oracle & Multimodal System & Biometric authentication & NIST Biometric Score set & Targeted integrity violation & Grey-box \\
 \hline
 \citet{laskov2014practical} & 2014 & Evasion & Support Vector Machine, Random Forest & PDF Malware Detection & Contagio and Operational & Confidence reduction & Black-box \\
 \hline
 \citet{nguyen2015deep} & 2015 & Evasion & Convolutional Neural Networks & Image Classification & ImageNet, MNIST & Misclassification & White-box \\
 \hline
  \citet{papernot2016limitations} & 2016 & Evasion & Convolutional Neural Network (LeNet) & Handwritten Digit Recognition & MNIST & Misclassification & White-box \\
 \hline
  \citet{moosavi2016deepfool} & 2016 & Evasion & Fully Connected Network, Convolutional Neural Network (LeNet, CaffeNet, GoogLeNet) & Handwritten Digit Recognition, Image Classification & MNIST, CIFAR-10, ImageNet & Misclassification & White-box \\
 \hline
   \citet{kurakin2016physical} & 2016 & Evasion & ImageNet Inception classifier & Image Classification & ImageNet & Misclassification & White-box \\
 \hline
   \citet{sharif2016accessorize} & 2016 & Evasion & VGG-Face Convolutional Neural Network & Facial Recognition System & PubFig Image  database & Targeted Misclassification, Misclassification & White-box \\
 \hline
    \citet{grosse2016adversarial} & 2016 & Evasion & Feed Forward Neural Network & Malware Detection & DREBIN Android malware dataset & Misclassification & White-box \\
 \hline
   \citet{sarkar2017upset} & 2017 & Evasion & Convolutional Neural Network & Handwritten Digit Recognition, Image Classification & MNIST, CIFAR-10 & Targeted Misclassification & Black-box \\
 \hline
    \citet{carlini2017towards} & 2017 & Evasion & Convolutional Neural Network & Handwritten Digit Recognition, Image Classification & MNIST, CIFAR-10, ImageNet & Targeted Misclassification & White-box \\
 \hline
    \citet{baluja2017adversarial} & 2017 & Evasion & Convolutional Neural Network, Fully Connected Network & Handwritten Digit Recognition, Image Classification & MNIST, ImageNet & Targeted Misclassification & White-box \\
 \hline
     \citet{moosavi2017universal} & 2017 & Evasion &Convolutional Neural Network(CaffeNet, VGG-F, VGG-16,VGG-19,GoogLeNet, ResNet-152) &  Image Recognition & ILSVRC 2012  &  Misclassification & White-box \\
 \hline
     \citet{chen2017zoo} & 2017 & Evasion & Convolutional Neural Network (Inception-v3) &  Image Classification & MNIST, CIFAR-10, ImageNet  &  Misclassification, Targeted Misclassification & Black-box \\
 \hline
     \citet{adate2017understanding} & 2017 & Evasion & Convolutional Neural Network &  Image Classification & MNIST  &  Misclassification & White-box \\
 \hline
     \citet{madry2017towards} & 2018 & Evasion & Convolutional Neural Network &  Image Classification & MNIST, CIFAR-10  &  Misclassification & White-box, Black-box \\
 \hline
     \citet{samanta2017towards} & 2017 & Evasion & Convolutional Neural Network &  Sentiment Analysis, Gender Detection & IMDB movie review, Twitter tweets  &  Confidence Reduction & White-box \\
 \hline
      \citet{dang2017evading} & 2017 & Evasion & SVM, Random Forest, Decision tree &  PDF Malware Detection & Contagio  & Misclassification & Black-box \\
 \hline
       \citet{hu2017generating} & 2017 & Evasion & Random Forest, Logistic Regression, Decision Trees, Support vector Machines, Multi-layer Perceptron (MLP) & Malware Detection & Dataset crawled from malwr.com  & Confidence reduction & Black-box \\
 \hline
     \citet{ilyas2018black} & 2018 & Evasion & Convolutional Neural Network &  Image Classification & ImageNet  &  Targeted Misclassification & Black-box \\
 \hline
      \citet{pengcheng2018query} & 2018 & Evasion & Convolutional Neural Network &  Image Classification & MNIST, CIFAR-10  & Misclassification & Black-box \\
 \hline
      \citet{dong2018boosting} & 2018 & Evasion & Convolutional Neural Network &  Image Classification & ImageNet  & Misclassification & White-box, Black-box \\
 \hline
     \citet{he2018decision} & 2018 & Evasion & Convolutional Neural Network &  Image Classification & MNIST, CIFAR-10  & Misclassification & White-box \\
 \hline
     \citet{chen2018ead} & 2018 & Evasion & Convolutional Neural Network &  Image Classification & MNIST, CIFAR-10, ImageNet  &  Targeted Misclassification & White-box \\
 \hline
     \citet{carlini2018audio} & 2018 & Evasion & Recurrent Neural Network (DeepSpeech) &  Automatic Speech Recognition & Mozilla Common Voice dataset &  Targeted Misclassification & White-box \\
 \hline
     \citet{eykholt2018robust} & 2018 & Evasion & Convolutional Neural Networks &  Stop Sign Recognition & LISA, GTSRB &  Targeted Misclassification & White-box \\
 \hline 
     \citet{chen2018attacking} & 2018 & Evasion & Convolutional Neural Networks (Inception-v3) &  Image Captioning & Microsoft COCO &  Targeted Misclassification & White-box \\
 \hline
     \citet{su2019one} & 2019 & Evasion & AllConv, NiN, VGG16 network &  Image Classification & CIFAR-10, ImageNet & Misclassification & Black-box \\
 \hline 
\end{tabularx}
\label{table:1.2}
\end{table*}


\subsection{AI Attack Surface}
The attack surface of an AI system is the total sum of vulnerabilities the AI model is exposed to during both the training and testing phases. It can also be described as a list of inputs that an adversary can use to attempt an attack on the system. As shown in Fig.\ref{analysis_framework}, the attack surface of an AI system can be viewed in terms of a generalized data processing pipeline which comprises of the training/test input data or objects, the learning algorithm/model, and the output data. At the testing stage, the input features are processed by the machine learning model to produce the class probabilities, which are further communicated to an external system in the form of an action to be acted upon. An adversary can attempt to attack this system by poisoning the training data, corrupting the learning model or tampering with the class probabilities.


\subsection{Attacker's Goals}
Adversarial goals can be broadly described based on the security properties of an information system, namely: \textit{confidentiality}, \textit{integrity}, \textit{availability} and \textit{privacy}. An adversary's goal in attacking the confidentiality of an AI system is to gather insights about the internals of the learning model or dataset and to use this information to carry out more advanced attacks. In other words, an attack on confidentiality can be targeted towards a model and its parameters or the training data \cite{papernot2018security}.  The goal of an attack on integrity is to modify the AI logic, through interaction with the system either at the learning or inference stage, while also controlling the model's outputs. In the example of a spam filter, an adversary can poison some data in the training dataset with the goal of changing the classification boundary such that a legitimate email is instead classified as spam and the adversary can evade detection without compromising normal system operations. According to \citet{papernot2016limitations}, four goals that impacts the integrity of a deep learning system are: (1) confidence reduction, (2) misclassification (3) targeted misclassification (4) source/target misclassification.

In adversarial settings, the availability of an AI solution can be attacked with the goal of disabling the system's functionality. For instance, an AI system can be flooded with incorrectly classified objects with the aim of causing the model to be unreliable or inconsistent in the target environment \cite{papernot2018security}. Attacks on the availability of a machine learning system could lead to many classification errors, making the system effectively unusable. Attacks on confidentiality and privacy are related in goal and method. An adversary seeks to violate the privacy of a machine learning system with the goal of causing it to reveal information about the training data and model. Potential privacy risks may involve an adversary with partial information about a training sample, attempting to manipulate a model into revealing information about unknown portions of the sample \cite{huang2011adversarial} or the extraction of the training dataset using the model's predictions.

\subsection{Attacker's Knowledge and Capabilities}
An attacker's knowledge of a machine learning system can be defined based on the knowledge of single components involved in the design of the system \cite{biggio2014security, biggio2013security}. An attacker can have different levels of knowledge of the system such as training data \begin{math} \mathcal{D} \end{math}, features \begin{math} \mathcal{X} \end{math}, learning algorithm \begin{math} f \end{math}, objective function \begin{math} \mathcal{L} \end{math}, and parameters \begin{math} \textbf{w} \end{math}. The adversary's knowledge can therefore be characterised as a vector \begin{math} \theta \in \Theta \end{math} where \begin{math} \theta = (\mathcal{D},\mathcal{X},f,\textbf{w}) \end{math}. On the other hand, adversarial capabilities refers to the level of information or knowledge about the machine learning system that is available to the adversary \cite{chakraborty2018adversarial}. For instance, in the example of a spam detector pipeline described earlier in this section, an adversary with access to or knowledge of the spam detector model used for classification has 'better' adversarial capabilities on the system relative to an adversary with only access to the 'tokenized' text data or incoming email.

\begin{figure}[h]
\centering\includegraphics[width=0.7\linewidth]{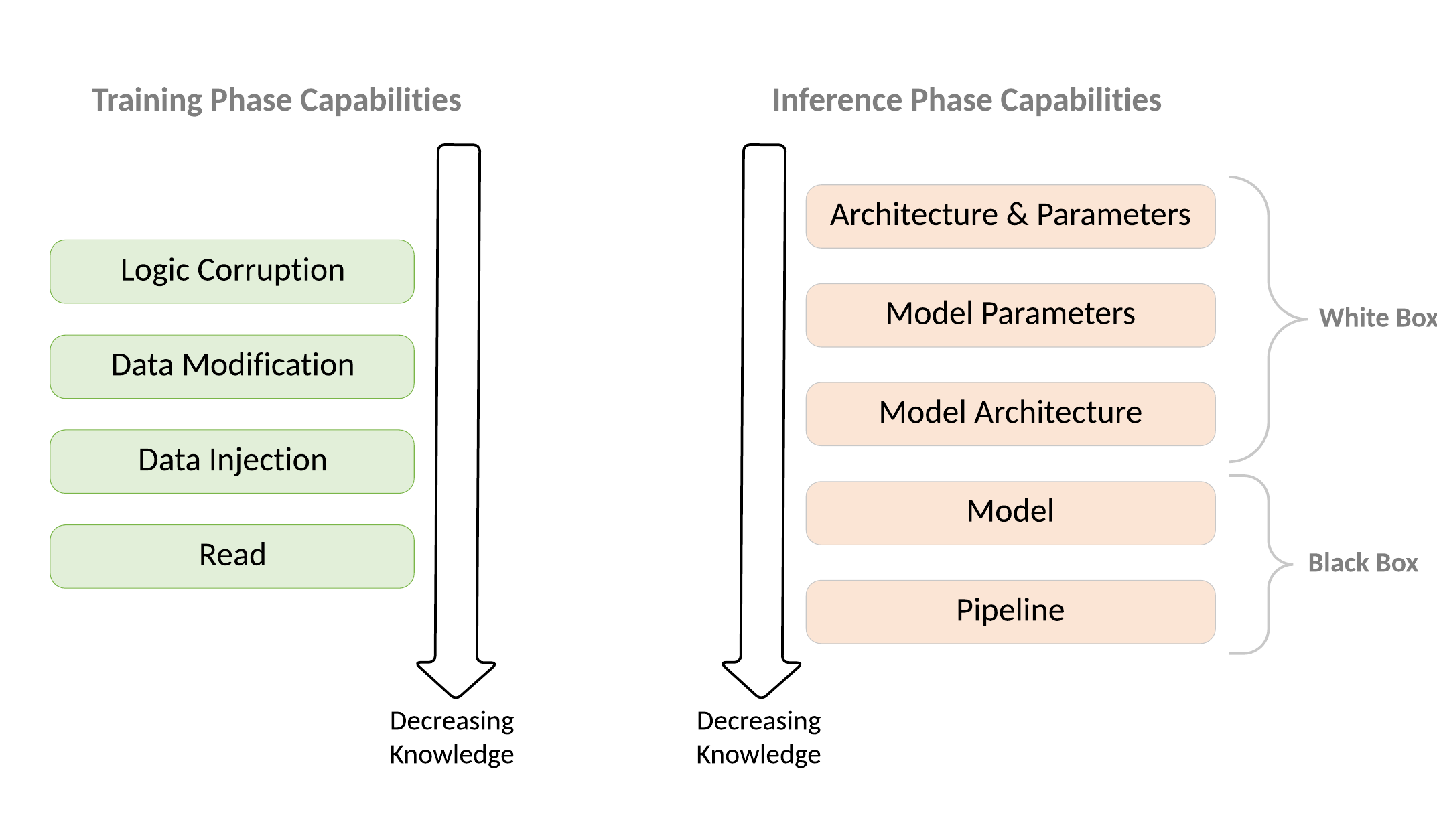}
\caption{Attacker's Capabilities}
\label{adversarial_capabilities}
\end{figure}

According to \citet{papernot2018security}, the adversarial capabilities in machine learning systems can be classified based on how they relate to the training and testing phases, as explained below. Fig. \ref{adversarial_capabilities} shows a relationship between an adversary's knowledge of a machine learning system and the adversary's capabilities.

\paragraph{Training Phase Capabilities}-- Attacks during the training phase seek to learn, influence or alter the performance of the model. The most straightforward attack on the training phase is one in which the adversary attempts to \textbf{read} or access a portion or all of the training data. An adversary that does not have access to the training data nor the learning algorithm can carry out a \textbf{data injection} attack by adding adversarial data to the existing training dataset. An adversary with knowledge of the training dataset but not the learning algorithm can directly poison the data or label (in supervised learning datasets) before it is used to train the target model. This type of attack capability is referred to as \textbf {data/label modification}. Lastly, an adversary with knowledge about the internals of an algorithm can carry out a \textbf{logic corruption} attack by altering the learning logic. Logic corruption attacks are the most advanced attacks on the training phase of a machine learning system and are difficult to defend against.

\paragraph{Testing Phase Capabilities}-- Testing phase attacks are referred to as \textbf{exploratory} attacks and according to \citet{barreno2006}, these attacks do not alter the training process nor influence the learning, but rather attempt to discover information about the state of the learner. Inference attacks rely on information about the model and its use in the target environment. As illustrated in Fig. \ref{adversarial_capabilities}, adversarial capabilities at the testing phase can be broadly classified into either black-box or white-box attacks. In a \textit{White-box} attack setting, the adversary has knowledge of everything about the model, including its training data, architecture, parameters, intermediate computations at hidden layers, as well as any hyper-parameters used for predictions \cite{biggio2018wild}, so we characterize white-box adversary's knowledge \begin{math} \theta_{Wb} = (\mathcal{D},\mathcal{X},f,\textbf{w}) \end{math}.\textit{Black-box} attacks assumes that the adversary has no knowledge about the model but is able to use information about previous input/output pairs to infer vulnerabilities in the model \cite{papernot2018security}. The model parameters and architecture are not accessible to the adversary as in the case of Machine Learning as a Service (MLaaS) platforms \cite{nasr2019comprehensive}, so \begin{math} \theta_{Bb} = (\hat{\mathcal{D}},\hat{\mathcal{X}},\hat{f},\hat{\textbf{w}}) \end{math}.

\subsection{Attack Strategy}
An attack strategy can be described as the method used by an adversary to modify the training and test dataset in other to optimize an attack \cite{wang2019security}. Thus, given an adversary's knowledge \begin{math} \theta \in \Theta \end{math} and a set of adversarial examples \begin{math} \mathcal{D'}_c \in \Phi(\mathcal{D}_c) \end{math}, an adversary's goal can be expressed in terms of an objective function \begin{math} \mathcal{A}(\mathcal{D'}_c,\theta) \in \mathbb{R} \end{math} which is a measure of the effectiveness of an attack with samples \begin{math} \mathcal{D'}_c \end{math}. An adversary's optimal attack strategy can therefore be expressed as:

\begin{equation}
\label{optimal_attack_strategy}
\mathcal{D}_c^* \in \operatorname*{argmax}_{\mathcal{D'}_c \in \Phi(\mathcal{D}_c)} \mathcal{A}(\mathcal{D'}_c, \theta)
\end{equation}

\subsection{Adversarial Attack Types}
The most common attack types against AI systems are discussed in this section based on the techniques used by an attacker.

\subsubsection{Poisoning Attacks} Poisoning attacks, sometimes referred to as causative attacks, are staged at the training phase where the adversary alters the training dataset by editing, injecting or removing samples with the aim of changing the decision boundary of the target model \cite{papernot2018security}. They can target the integrity of adaptive or online classifiers, as well as classifiers being retrained on data collected at test time \cite{biggio2014security}. Some examples of poisoning attacks on AI systems are presented in Table \ref{table:1.1}. According to \citet{munoz2019}, many of the data used in training machine learning systems come from untrusted sources, so the problem of poisoning attacks can be described as related to the reliability of large amount of data collected by these systems. While an adversary may be unable to directly access an existing training dataset, the ability to provide new training data via web-based repositories and honeypots provides an opportunity to poison training data \cite{biggio2012poisoning}.

Some literature \cite{biggio2014security, biggio2018wild, munoz2019} have described two main scenarios of poisoning attacks against multi-class classification systems namely:

\textit{Error-generic poisoning attacks} - These are the most common poisoning attacks in which the attackers aim to cause a denial of service attack in the system, by producing as many misclassifications on targeted data points as possible irrespective of the classes in which they occur. Using formulation of the attack strategy in Eq.\ref{optimal_attack_strategy}, the error-generic poisoning attack can be expressed as:

\begin{equation}
\label{error_generic_poisoning}
\mathcal{D}_p^* \in \operatorname*{argmax}_{\mathcal{D}_p \in \Phi(\mathcal{D}_p)} \mathcal{A}(\mathcal{D}_p, \theta) = \mathcal{L}(\mathcal{D}_{target},\textbf{w}(\mathcal{D}_p))
\end{equation}

where the objective function an adversary aims to maximize, \begin{math} \mathcal{A} \end{math}, is defined in terms of loss function, \begin{math} \mathcal{L} \end{math}, computed on set of points, \begin{math} \mathcal{D}_{target} \end{math}, targeted by the attacker. The loss function, \begin{math} \mathcal{L} \end{math}, is a function of parameters, \begin{math} \textbf{w} \end{math}, which depends on injected poisoning points, \begin{math} \mathcal{D}_p \end{math}.

\textit{Error-specific poisoning attacks} - The aim of an adversary in error-specific poisoning attacks is to cause specific misclassifications \cite{munoz2019}. The attack strategy in this case can be expressed as:

\begin{equation}
\label{error_specific_poisoning_1}
        \mathcal{D}_p^* \in \operatorname*{argmax}_{\mathcal{D}_p \in \Phi(\mathcal{D}_p)} \mathcal{A}(\mathcal{D}_p, \theta) =                 \operatorname*{argmax}_{\mathcal{D}_p \in \Phi(\mathcal{D}_p)} - \mathcal{L}(\mathcal{D'}_{target},\textbf{w}(\mathcal{D}_p))
\end{equation}

where \begin{math} \mathcal{D}_{target} \end{math} contains the same samples as \begin{math} \mathcal{D'}_{target} \end{math} but with labels chosen by the adversary based on desired misclassifications. The loss function \begin{math} \mathcal{L} \end{math} is expressed with negative sign because the adversary aims to minimize loss on the desired labels. Therefore, Eq.\ref{error_specific_poisoning_1} can equivalently be expressed as:

\begin{equation}
\label{error_specific_poisoning_2}
\mathcal{D}_p^* \in \operatorname*{argmin}_{\mathcal{D}_p \in \Phi(\mathcal{D}_p)} \mathcal{L}(\mathcal{D'}_{target},\textbf{w}(\mathcal{D}_p))
\end{equation}

\subsubsection{Oracle Attacks} 
\begin{figure}[h]
\centering\includegraphics[width=0.6\linewidth]{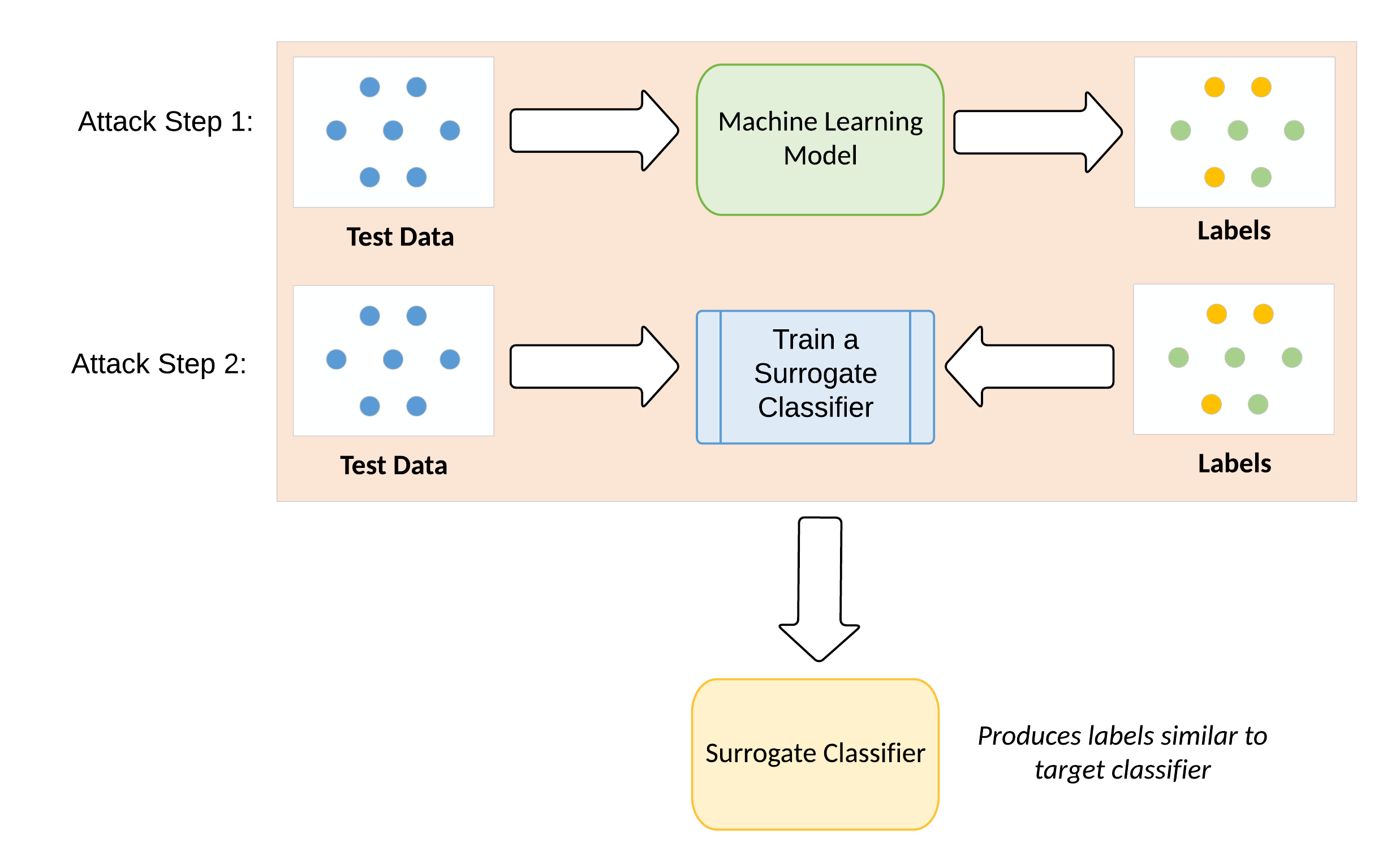}
\caption{Steps of an Oracle Attack}
\label{oracle_attack_steps}
\end{figure}

These are exploratory attacks where an adversary uses samples to collect and infer information about a model or its training data \cite{tabassi2019taxonomy}. Many classifiers are considered the intellectual property of the organizations they belong to; for instance in Machine Learning as a Service (MLaaS) platforms such as Microsoft Machine Learning, Amazon Machine Learning and Google Prediction API. In these environments, simple APIs are made available for the customer to upload data and for training and querying models \cite{shokri2017membership}. An adversary can perform an oracle attack against such models by using an API to feed inputs into the learning model in order to observe the model's outputs. An adversary with no knowledge of a classifier can use input-output pairings obtained from oracle attacks to train a surrogate model that operates much like the target model \cite{shi2017evasion}. The steps an adversary uses to carry out an oracle attack are described in Fig. \ref{oracle_attack_steps}. Without prior knowledge of the original classifier and class information of each sample, the adversary can only feed some test data into the classifier and collect the resulting labels which are then used to train a surrogate classifier, that is, a functionally equivalent classifier. The surrogate classifier can then be used to generate adversarial examples for use in evasion attacks against the target model. In oracle attacks, sometimes referred to as exploratory attacks, an adversary uses samples to collect and infer information about a model or its training data. Oracle attack types include membership inference attacks, inversion attacks and extraction attacks \cite{tabassi2019taxonomy}. 

Many machine learning models are considered the intellectual property of the organizations that created them. For instance, Microsoft Machine Learning, Amazon Machine Learning and Google Prediction API are all Machine Learning as a Service (MLaaS) platforms and are owned and operated by Microsoft, Amazon and Google respectively. In these environments, simple APIs are made available for customers to upload data, and for training and querying models \cite{shokri2017membership}. An adversary can perform an oracle attack against such models by using an API to feed inputs into the learning model in order to observe the model's outputs. An adversary with no knowledge of a target model can use input-output pairings obtained from oracle attacks to train a surrogate model which is functionally equivalent to the target model \cite{shi2017evasion}. The surrogate model can then be used to generate adversarial examples for use in evasion attacks against the target model. Oracle attacks include membership inference attacks, inversion attacks and extraction attacks. 

\textit{Membership Inference Attacks} - In membership inference attacks, the adversary seeks to determine if a given data point belongs to the training dataset analyzed to learn the model's parameters \cite{papernot2018security}. For instance, an adversary can violate the privacy of a clinical patient by using membership inference attack to know if the patient's health record was part of the dataset used in training a model that diagnoses a particular disease. \citet{shokri2017membership} demonstrated how the membership inference attacks can be carried out against black box models by exploiting the difference in the target model's behaviour on data 'seen' during training versus data seen for the first time.

\textit{Inversion Attacks} - Model inversion attacks, referred to as training data extraction attacks in some literature, enables an adversary to reconstruct training inputs from a model's predictions \cite{fredrikson2015model}. The extracted inputs are not specific points but rather an average representation of data that belong to a class. \citet{fredrikson2014privacy} introduced inversion attacks by considering a case where an adversary is able to infer a patient's genetic marker from the predictions of a linear model designed to predict a stable dosage of the drug Warfarin. This approach clearly illustrates privacy issues relating to providing API access to ML models trained on sensitive data \cite{papernot2018security}.

\textit{Extraction Attacks} - Model extraction attacks involves an adversary who uses observed input-output pairs, \begin{math} (x_i,y_i) \end{math}, from the predictions of some target model, \begin{math} f \end{math}, to extract a set of parameters and attempts to learn a model, \begin{math} \hat{f} \end{math}, that closely approximates \begin{math} f \end{math}. \citet{tramer2016stealing} demonstrated extraction attacks against two ML-as-a-Service models via their exposed APIs. While extraction attacks could subvert model monetization, it also violates training data privacy and facilitates evasion attacks.

\subsubsection{Evasion Attacks} Evasion attacks are the most common type of attacks on AI systems \cite{chakraborty2018adversarial}. An adversary directly manipulates the input samples at test time to avoid detection \cite{biggio2013security}. While the evasion attack scenarios explored in many literature have mainly focused on machine learning classification tasks, deep neural networks have been found to be highly vulnerable to this kind of threat \cite{munoz2019, biggio2013evasion, szegedy2013intriguing, papernot2016transferability}. However, deep networks also have the advantage of being able to represent functions that can resist adversarial perturbations unlike other shallow linear models \cite{goodfellow2014explaining}. Evasion attacks do not alter the behaviour of a system but instead exploit vulnerabilities in the system using adversarial examples to produce the desired errors. The approach used in evasion attacks is based on the rationale that without prior knowledge of a classifier's decision function, an adversary can learn a surrogate classifier using a surrogate data to reliably evade the targeted classifier. Adversarial samples are often perceptually indistinguishable from the 'clean' samples but yet pose serious security threats for machine learning applications \cite{kurakin2016adversarial}. Some examples of evasion attacks on AI system are presented in Table \ref{table:1.2}.

\citet{melis2017deep} described two possible evasion attack settings based on multi-class classifiers namely: Error-generic evasion attacks and Error-specific evasion attacks.
Many evasion attack scenarios can be derived from these attack categories.

\textit{Error-generic evasion attacks} - These refers to attacks where the adversary is simply interested in misleading classification at test time irrespective of the output class predicted by the classifier. In multi-class classifiers, the predicted class \begin{math} c^* \end{math} is a class for which the discriminant function, \begin{math} f_k(x) \end{math}, for a given input sample, \begin{math} x \end{math}, is maximum:

\begin{equation}
\label{predicted_class}
c^* = \operatorname*{argmax}_{k=1,...,c}f_k(x)
\end{equation}

The error-generic evasion attack problem can be formulated mathematically as:

\begin{equation}
\label{error_generic_evasion}
\begin{split}
x_e^* \in \operatorname*{argmin}_{x_e \in \Phi(x_e)} f_k(x_e) - \operatorname*{max}_{j \neq k} f_j(x_e), \; \; \textrm{s.t.} \; d(x,x_e) \leqslant d_{max}
\end{split}
\end{equation}

where \begin{math} f_k(x_e) \end{math} is the discriminant function associated with the true class \begin{math} k \end{math} of the source sample \begin{math} x \end{math} and \begin{math} \operatorname*{max}_{j \neq k} f_j(x_e) \end{math} is the discriminant function of the closest competing incorrect class. The optimization problem constraint \begin{math} d(x,x_e) \leqslant d_{max} \end{math} limits the perturbation, \begin{math} d_{max} \end{math} between the original sample, \begin{math} x \end{math}, and the adversarial sample, \begin{math} x_e \end{math}, point given in terms of distance in the input space \cite{munoz2019}.

\textit{Error-specific evasion attacks} - In this evasion attack setting, the adversary is interested in misleading classification with the aim of producing a specific type of error. The problem of error-specific evasion attacks is formulated similarly as the error-generic evasion attack with the exception that the objective function is maximized and \begin{math} f_{k_e} \end{math} denotes the discriminant function associated with the class \begin{math} k_e \end{math} which the adversarial example should be assigned \cite{melis2017deep}:

\begin{equation}
\label{error_specific_evasion}
\begin{split}
x_e^* \in \operatorname*{argmin}_{x_e \in \Phi(x_e)} \; \Big(\operatorname*{max}_{j \neq k_e} f_j(x_e)\Big) -  f_{k_e}(x_e), \;\;\textrm{s.t.}\; d(x,x_e) \leqslant d_{max}
\end{split}
\end{equation}

\subsection {Adversarial Examples: Crafting Methods}
This is an important research direction as it provides guidance on vulnerabilities in AI systems which could potentially be exploited by an adversary. In this section, we review some of the methods for generating adversarial examples in literature.

\subsubsection{Box-Constrained L-BFGS}
\citet{szegedy2013intriguing} made a groundbreaking discovery on the stability of neural network networks to small imperceptible input perturbations and showed how these perturbations could arbitrarily change the network's prediction. They formalized the problem of finding the optimal perturbation in terms of the L2 norm. To find the optimal perturbation \begin{math} \eta \end{math} for a given clean input image \begin{math} \mathbf{X} \in \mathbb{R}^m \end{math} and target label \begin{math} l \in \{1,...,k\} \end{math}, the aim is to solve a box-constrained optimization problem:

\begin{equation}
\label{l_bfgs}
\operatorname*{min}_\eta {\left\|\eta\right\|}_2 \;\; s.t. \; F(\mathbf{X} + \eta) = l; \;\; \mathbf{X} + \eta \in [0,1]^m
\end{equation}
where \begin{math} F : \mathbb{R}^m \to \{1...k\} \end{math} is the deep neural network classifier which is assumed to have an associated continuous loss function \begin{math} \mathcal{L}: \mathbb{R}^m \times \{1...k\} \to \mathbb{R}^+ \end{math}. The problem of Eq.\ref{l_bfgs} is difficult to solve, so \citeauthor{szegedy2013intriguing} proposed an approximation using the box-constrained L-BFGS:

\begin{equation}
\label{l_bfgs_2}
\operatorname*{min}_\eta c{\|\eta\|} + \mathcal{L}(\mathbf{X} + \eta,l) \;\; s.t. \;\; \mathbf{X} + \eta \in [0,1]^m
\end{equation}
An approximate solution to Eq. \ref{l_bfgs_2} is obtained by performing a line-search to find the minimum constant \begin{math} c > 0 \end{math} for which the minimizer \begin{math} \eta \end{math} satisfies \begin{math} F(\mathbf{X}+\eta) = l \end{math}. According to \citet{moosavi2016deepfool}, this method of generating adversarial perturbations is less efficient and does not scale to large datasets.

\subsubsection{Fast Gradient Sign Method (FGSM)} \citet{goodfellow2014explaining} introduced FGSM, an efficient method for generating adversarial examples in the computer vision context. Their work established that the linear behaviour of deep neural network is the main reason they are vulnerable to adversarial examples. The Fast Gradient Sign Method is based on linearizing the cost function of a deep neural network and solving for the perturbation \begin{math} \eta \end{math} that maximizes the \begin{math} L_\infty \end{math} norm. In closed form:

\begin{equation}
 \eta = \epsilon sign(\bigtriangledown_{\mathbf{X}} J(\mathbf{X}, y))
\end{equation}

\begin{equation}
 \tilde{\mathbf{X}} = \mathbf{X} + \eta
\end{equation}
where \begin{math} \epsilon \end{math} is a hyper-parameter to be selected by the adversary, \begin{math} \mathbf{X} \end{math} represents the input to the model, \begin{math} y \end{math} represents the targets associated with \begin{math} \mathbf{X} \end{math} and \begin{math} \bigtriangledown_{\mathbf{X}} J(\mathbf{X}, y) \end{math} is the gradient of the loss function  w.r.t the input \begin{math} \mathbf{X} \end{math}, and can be computed via back-propagation. According to \citet{kurakin2016adversarial}, this method is simple, fast and computationally efficient as it does not require iterative procedure to compute the adversarial examples. Nevertheless, the FGSM method only provides a coarse approximation of the optimal perturbation vectors \cite{moosavi2016deepfool}.

\subsubsection{Basic Iterative Method (BIM)} The Basic Iterative Method (BIM) is simply an extension of the FGSM and was introduced by \citet{kurakin2016physical}. It is an iterative method for crafting adversarial examples by applying the FGSM technique multiple times with small step size:

\begin{equation}
    \begin{split}
        \tilde{\mathbf{X}}_0 = \mathbf{X}, \;\;\; \tilde{\mathbf{X}}_{N+1} = \textrm{Clip}_{\mathbf{X},\epsilon} \bigg\{\tilde{\mathbf{X}}_N +
        \alpha sign (\bigtriangledown_{\mathbf{X}} J(\tilde{\mathbf{X}}_N, y)) \bigg\}
    \end{split}
\end{equation}

\subsubsection{Jacobian-based Saliency Map Attack (JSMA)} The JSMA was introduced by  \citet{papernot2016limitations} as a class of algorithms which an adversary can use to reliably generate adversarial examples based on a precise understanding of the input features of a deep neural network that most significantly impacts the output. This approach is distinct from previous approaches for generating adversarial examples because rather than using the gradient of the DNN's cost function to compute the perturbation vector, it instead computes the forward derivative of the neural network directly and then uses the concept of adversarial saliency maps to highlight regions of the input domain that leads to significant changes in the network outputs. The forward derivative \begin{math} J_{\mathbf{F}}(\mathbf{X}) \end{math} for a given sample \begin{math} \mathbf{X} \end{math}, also referred to as the Jacobian of the multidimensional function \begin{math} \mathbf{F} \end{math} learned by the DNN during training is expressed as:

\begin{equation}
\label{jsma}
J_{\mathbf{F}}(\mathbf{X}) = \frac{\partial\mathbf{F(X)}}{\partial\mathbf{X}} = \left[ \frac{\partial\mathbf{F}_j(\mathbf{X})}{\partial x_i} \right]_{i \in 1...M, j \in 1...N}
\end{equation}

Based on the experimental evaluations carried out in their work on LeNet, a well-studied deep neural network used for handwritten digit recognition,  \citet{papernot2016limitations} demonstrated that the JSMA algorithms can reliably produce adversarial examples which can cause misclassifications in specific target DNN models. 

\subsubsection{Iterative Least-Likely Class Method (ILLC)}
The iterative least-likely class method (ILLC), introduced by \citet{kurakin2016physical}, is another technique that has been demonstrated in literature for generating adversarial examples for targeted attacks against deep neural networks in the computer vision context. In contrast to FGSM and BIM, this method enables an adversary to create adversarial examples that will be classified by a DNN as a specific desired target class. The least-likely class according to the prediction of a trained DNN on image \begin{math} \mathbf{X} \end{math} is usually very different from the true class and can be expressed as shown in Eq. \ref{least_likely_class}.

\begin{equation}
\label{least_likely_class}
y_{LL} = \operatorname*{argmin}_y \left\{p(y|\mathbf{X})\right\}
\end{equation}

According to \citet{kurakin2016adversarial}, the procedure for generating an adversarial image which is classified as \begin{math} y_{LL} \end{math} is:

\begin{equation}
\label{least_likely_procedure}
    \begin{split}
        \tilde{\mathbf{X}}_0 = \mathbf{X}, \;\;\; \tilde{\mathbf{X}}_{N+1} = Clip_{\mathbf{X},\epsilon} \bigg\{\tilde{\mathbf{X}}_N - 
        \alpha sign (\bigtriangledown_{\mathbf{X}} J(\tilde{\mathbf{X}}_N, y_{LL})) \bigg\}
    \end{split}
\end{equation}

\subsubsection{Universal Perturbations for Steering to Exact Targets (UPSET)}
Another method for generating adversarial images that looks similar to the input but fools classifiers is the method known as universal perturbations for steering to exact targets (UPSET) proposed by \citet{sarkar2017upset}. Given an \begin{math} n \end{math} class setting, UPSET seeks to produce \begin{math} n \end{math} universal perturbations \begin{math} \eta_j \end{math}, \begin{math} j \in \{1,2,...,n\} \end{math} such that when the perturbation is added to any image that does not belong to target class \begin{math} j \end{math}, the perturbed image is classified as being from target class \begin{math} j \end{math}. UPSET uses a residual generating network \begin{math} R \end{math} which uses target class \begin{math} t \end{math} and outputs a perturbation \begin{math} \eta_t = R(t) \end{math} which is of similar dimension to input image \begin{math} \mathbf{X} \end{math}. The adversarial image \begin{math} \tilde{\mathbf{X}} \end{math} is expressed as:

\begin{equation}
\label{upset}
\tilde{\mathbf{X}} = \operatorname*{max}(\operatorname*{min}(s \times R(t) + \mathbf{X},1),-1),
\end{equation}
where \begin{math} s \end{math} is a scaling factor and the pixel values in \begin{math} \mathbf{X} \end{math} are normalized to be within [-1, 1].

\subsubsection{Antagonistic Network for Generating Rogue Images (ANGRI)}
The Antagonistic Network for Generating Rogue Images (ANGRI) is another method for generating adversarial images proposed by \citet{sarkar2017upset}. As opposed to UPSET which produces image-agnostic perturbations, ANGRI instead produces image-specific perturbations as its output depends on the input image. Given an input image, \begin{math} \mathbf{X} \end{math}, belonging to class \begin{math} C_\mathbf{X} \end{math} and a target class, \begin{math} t \neq C_\mathbf{X} \end{math}, ANGRI generates an adversarial image, \begin{math} \tilde{\mathbf{X}} \end{math}, which the classifier classifies as an image from target class \begin{math} t \end{math}.

\subsubsection{DeepFool}
DeepFool, introduced by \citet{moosavi2016deepfool} is an iterative method for computing adversarial examples that fools deep neural networks. In comparison to both the box-constrained L-BFGS and FGSM methods discussed earlier, DeepFool is more computationally efficient and generates adversarial perturbations that are more imperceptible. The DeepFool method is based on the assumption that neural networks are linear with each class separated by hyperplanes \cite{carlini2017towards}. Therefore the minimum perturbation of a linearized binary classifier is computed as:

\begin{equation}
\label{deepfool}
\operatorname*{argmin}_{\eta_i}\|\eta_i\|_2 \;\; s.t. \;\; f(x_i) + \bigtriangledown f(x_i)^T \eta_i = 0.
\end{equation}
where \begin{math} \eta_i \end{math} is the perturbation at iteration \begin{math} i \end{math}, \begin{math} x \end{math} is the input image and \begin{math} f \; : \; \mathbb{R}^n \to \mathbb{R} \end{math} is an arbitrary image classification function. \citeauthor{moosavi2016deepfool} also extended DeepFool multi-class classifiers. Experimental results presented in their work from testing the DeepFool algorithm on CNN architectures using the MNIST, CIFAR-10 and ImageNet classification datasets show that method accurately estimates the minimum perturbation required to change classification labels of deep neural networks.

\subsubsection{Carlini \& Wagner}

\citet{carlini2017towards} introduced one of the most common and effective algorithms for crafting adversarial examples to date. Their approach is also based on the formulation of adversarial examples introduced by \citet{szegedy2013intriguing}. Given the problem of finding an adversarial example for input image \begin{math} \mathbf{X} \end{math}:

\begin{equation}
\label{carlini_1}
\begin{aligned}
\operatorname*{minimize} \;\; \mathcal{D}(\mathbf{X},\mathbf{X}+\delta)\\
\textrm{such that} \;\; \mathcal{C}(\mathbf{X}+\delta) =t,\\
\mathbf{X}+\delta \in [0,1]^n
\end{aligned}
\end{equation}

The goal therefore is to find a small perturbation \begin{math} \delta \end{math} that minimizes some distance metric \begin{math} \mathcal{D}(\mathbf{X},\mathbf{X}+\delta) \end{math}, usually specified in terms of \begin{math} L_p \end{math} norms, in order to cause a misclassification. Constraint \begin{math} \mathcal{C}(\mathbf{X}+\delta) =t \end{math} ensures that the input image is mis-classified and is highly non-linear. This constraint makes the optimization problem of Eq.\ref{carlini_1} difficult to solve and while \citet{szegedy2013intriguing} solved this problem using the L-BFGS technique, \citeauthor{carlini2017towards} solved the problem by re-formulating it and expressing the constraint as an objective function \begin{math} f \end{math} such that when \begin{math} \mathcal{C}(\mathbf{X}+\delta) =t \end{math} is satisfied, \begin{math} f(\mathbf{X}+\delta) \leq 0 \end{math} is also satisfied. After instantiating distance metric \begin{math} \mathcal{D} \end{math} with an \begin{math} L_p \end{math} norm, the \citeauthor{carlini2017towards} formulated optimization problem becomes:

\begin{equation}
\label{carlini_2}
\begin{aligned}
\operatorname*{minimize} \;\; \left \|\delta \right \|_p + c \cdot f(\mathbf{X}+\delta)\\
\textrm{such that} \;\; \mathbf{X}+\delta \in [0,1]^n
\end{aligned}
\end{equation}
Constraint \begin{math} \mathbf{X}+\delta \in [0,1]^n \end{math}, expressed as a box-constraint ensures that the adversarial example is a valid image. To find adversarial examples that will have a low \begin{math} L_2 \end{math} metric distortion, \citeauthor{carlini2017towards} used a \textit{change of variable} method to optimize a new variable \begin{math} w \end{math} rather than variable \begin{math} \delta \end{math}, setting:

\begin{equation}
\label{carlini_3}
\delta_i = \frac{1}{2}( \tanh (w_i)+1)-\mathbf{x}_i
\end{equation}

\subsubsection{Houdini}
Houdini, introduced by \citet{cisse2017houdini}, is an method for reliably generating adversarial examples which are directly tailored for a given combinatory or non-differentiable task loss of interest. Different from many existing methods which are mainly applicable to image classification tasks, Houdini is a more flexible approach which has been successfully applied to a range of applications such as voice recognition, human pose estimation \cite{xiao2019advit} and semantic segmentation. According to \citeauthor{cisse2017houdini}, the problem of finding an adversarial example which can fool a neural network model \begin{math} f_\theta \end{math} with respect to task loss \begin{math} l(.) \end{math} for a chosen p-norm and noise parameter \begin{math} \epsilon \end{math} involves solving:

\begin{equation}
\label{houdini_1}
\tilde{x} = \operatorname*{argmax}_{\tilde{x}:{\|\tilde{x}-x\|}_p \leq \epsilon} l(y_\theta(\tilde{x}),y)
\end{equation}
In most cases, the task loss \begin{math} l(.) \end{math} is a combinatory quantity that is difficult to optimize. As such, it is replaced with a differentiable surrogate loss \begin{math} \bar{l}(y_\theta(\tilde{x}),y) \end{math}. For a given example \begin{math} (x,y) \end{math}, \citeauthor{cisse2017houdini} proposed a surrogate named Houdini and defined as:

\begin{equation}
\label{houdini_2}
    \begin{split}
        \bar{l}_H(\theta,x,y)=\mathbb{P}_{\gamma \sim \mathcal{N}(0,1)}\bigg[ f_\theta(x,y) - 
        f_\theta(x,\hat{y})<\gamma \bigg] \cdot l(\hat{y},y)
    \end{split}
\end{equation}
The first term of Eq. \ref{houdini_2} is the stochastic margin and reflects the confidence of the neural network model in its prediction while the second term is the task loss.

\subsubsection{Universal Adversarial Perturbations (UAP)}
Universal adversarial perturbations, introduced by \citet{moosavi2017universal}, is a systematic algorithm for generating quasi-imperceptible universal perturbations for deep neural networks. In their work, they demonstrated how the UAP algorithm can be used to generate a single perturbation vector that can be used to fool neural networks on most natural images. Different from the notion of adversarial perturbations presented in \cite{szegedy2013intriguing}, \citeauthor{moosavi2017universal} examined universal perturbations that are common to most data points belonging to a data distribution. Given a distribution of images \begin{math} \mu \in \mathbb{R}^d\end{math} and classification function \begin{math} \hat{k} \end{math} that outputs estimated label \begin{math} \hat{k}(x) \end{math} for each image input \begin{math} x \in \mathbb{R}^d \end{math}, the focus is to obtain universal perturbation vectors \begin{math} \upsilon \in \mathbb{R}^d \end{math} that can fool \begin{math} \hat{k} \end{math} on most data points sampled from \begin{math} \mu \end{math} such that:

\begin{equation}
\label{uap}
\hat{k}(x+\upsilon)\neq \hat{k}(x) \;\; for \;\; "most"\;\; x \sim \mu
\end{equation}

The UAP algorithm seeks a universal perturbation \begin{math} \upsilon \end{math} that satisfies the constraints:

\begin{equation}
\label{uap_constraint_1}
\left\| \upsilon \right\|_p \leq \xi,
\end{equation}

\begin{equation}
\label{uap_constraint_2}
\underset{x \sim \mu}{\mathbb{P}} \left( \hat{k}(x+\upsilon) \neq \hat{k}(x) \right) \geq 1- \delta
\end{equation}
where parameter \begin{math} \xi \end{math} controls the magnitude of \begin{math} \upsilon \end{math} and \begin{math} \delta \end{math} quantifies the target fooling rate for all samples images from distribution \begin{math} \mu \end{math}.

\subsubsection{ATNs}
An Adversarial Transformation Networks (ATN) is a feed-forward neural network introduced by \citet{baluja2017adversarial} for generating adversarial examples. The ATN is formally expressed as a neural network:

\begin{equation}
\label{atn}
g_{f,\theta}(\mathbf{x}): \mathbf{x} \in \mathcal{X} \to \mathbf{x}'
\end{equation}
where \begin{math} \theta \end{math} is the parameter vector of \begin{math} g \end{math}, \begin{math} f \end{math} is the target network or set of networks, and \begin{math} \mathbf{x}' \sim \mathbf{x} \end{math} but \begin{math} \operatorname*{argmax} f(\mathbf{x}) \neq \operatorname*{argmax} f(\mathbf{x}') \end{math}. Obtaining \begin{math} g_{f,\theta} \end{math} requires solving the optimization problem:

\begin{equation}
\label{atn_optimization}
    \begin{split}
        \operatorname*{argmin}_\theta \sum_{\mathbf{x}_i \in \mathcal{X}} \beta L_\mathcal{X}\left(g_{f,\theta}(\mathbf{x}_i),\mathbf{x}_i\right)\; +
        \;L_\mathcal{Y}\left(f(g_{f,\theta}(\mathbf{x}_i)),f(\mathbf{x}_i)\right)
    \end{split}
\end{equation}
where \begin{math} L_\mathcal{X} \end{math} is the loss function in the input space, \begin{math} L_\mathcal{Y} \end{math} is the loss function on the output space of \begin{math} f \end{math} and \begin{math} \beta \end{math} is the weight that balances both loss functions.
\citeauthor{baluja2017adversarial} presented two different approaches for generating adversarial examples using an ATN: \textit{(1) training the ATN to only generate the perturbation to \begin{math} \mathbf{x} \end{math}}, or \textit{(2) training the ATN to generate an adversarial autoencoding of \begin{math} \mathbf{x} \end{math}}. Experimental results show that ANTs can generate a wide range of adversarial examples targeting a single network. The authors argued that the unpredictable diversity of ANTs makes it a viable attack that can bypass many defenses, including those trained using adversarial examples from existing methods.

\subsubsection{Projected Gradient Descent (PGD)}
The PGD attack, proposed by \citet{madry2017towards}, is a white-box $l_\infty$ attack that generates adversarial examples by using the local first order information about the neural network. In contrast to other $l_\infty$-bounded attacks such as FGSM \cite{goodfellow2014explaining}, the PGD is a more powerful multi-step attack as it doesn't limit the amount of time and effort an adversary can invest into finding the best attack. The PGD attack algorithm is based on the same intuition as the FGSM which involves solving the saddle-point problem:

\begin{equation}
\label{pgd}
    \begin{aligned}
        \operatorname*{min}_{\theta} \mathbb{E}_{(x,y) \sim \mathcal{D}}\left[ \operatorname*{max}_{\delta \in S}L(\theta,x+\delta,y) \right]
    \end{aligned}
\end{equation}
where $\mathcal{D}$ is the data distribution over pairs of examples $x\in \mathbb{R}^d$ and corresponding labels $y \in [k]$, and $\theta$ is the set of model parameters.

\subsubsection{One Pixel}
\citet{su2019one} proposed a method for generating one-pixel adversarial perturbations, based on differential evolution, which can be used for low dimension black-box attacks against deep neural networks. In formalizing the one pixel method, \citeauthor{su2019one} represented an input image by an \begin{math}n \end{math}-dimensional vector \begin{math} \mathbf{x}=(x_1,...,x_n) \end{math}, with each scalar elements also corresponding to one pixel. Given a target image classifier \begin{math} f \end{math} with original input image \begin{math} \mathbf{x}\end{math} correctly classified as class \begin{math} t \end{math}, the probability of \begin{math} \mathbf{x} \end{math} being classified as belonging to \begin{math} t \end{math} is \begin{math} f_t(\mathbf{x}) \end{math}. The vector \begin{math} r(\mathbf{x}) = (r_1,...,r_n)   \end{math} is an additive adversarial perturbation with respect to \begin{math} \mathbf{x} \end{math}. The goal therefore is to find the optimized solution \begin{math} r(\mathbf{x})^* \end{math}:

\begin{equation}
\label{one_pixel}
\begin{aligned}
\operatorname*{maximize}_{r(\mathbf{x})^*} \;\;\; f_{adv}(\mathbf{x}+r(\mathbf{x})) \;\;
\textrm{such that} \;\; \left\| r(\mathbf{x})\right\|_0 \leq d
\end{aligned}
\end{equation}
where \begin{math} d \end{math} is a small number and in the case of one-pixel attack, \begin{math} d=1
\end{math}.

While one pixel has only so far been demonstrated in deep neural networks for image classification tasks, the authors argue that it can potentially be extended to other deep learning applications such as natural language processing and speech recognition in future works.

\subsubsection{OPTMARGIN}
\citet{he2018decision} introduced the OPTMARGIN attack which can be used to generate low-distortion adversarial examples that are robust to small perturbations. \citeauthor{he2018decision} demonstrated in their work that adversarial examples generated using the OPTMARGIN attack can successfully evade the region-based classification defense method introduced in \cite{cao2017mitigating}, as a robust defense against low-distortion adversarial examples. The OPTMARGIN attack involves creating a surrogate model \begin{math} f_i(x) \end{math}, which is functionally equivalent to the region-based classifier \cite{cao2017mitigating}, and uses the surrogate model to classify a small number \begin{math} n \end{math} of perturbed data points, such that \begin{math} f_i(x) = f(x+v_i) \end{math}; where \begin{math} i=1,...,n \end{math}, \begin{math} f \end{math} is the point classifier used in the region-based classifier and \begin{math} v_i \end{math} are perturbations applied to input \begin{math} x \end{math}. If \begin{math} Z(x) \end{math} is the \begin{math} |C| \end{math}-dimensional vector of class weights that \begin{math} f \end{math} internally uses to classify \begin{math} x \end{math}, then the loss function for the surrogate model with data point \begin{math} i \end{math} can be expressed as:

\begin{equation}
\label{optmargin}
    \begin{split}
        l_i(x')=l(x' + v_i) = \operatorname*{max}\bigg(-\kappa, Z(x'+v_i)_y \; - \;
        \operatorname*{max} \left\{Z(x'+v_i)_j : j \neq y\right\}\bigg)
    \end{split}
\end{equation}
where \begin{math} \kappa \end{math} is the confidence margin. In the OPTMARGIN attack, \begin{math} \kappa \end{math} is set to \begin{math} 0 \end{math}, which implies that the model simply misclassifies its input. As an extension to the Carlini \& Wagner attacks \cite{carlini2017towards}, the OPTMARGIN's objective function uses the sum of the loss function for each perturbed data point, resulting in the minimization problem:

\begin{equation}
\label{optmargin_minimization}
\operatorname*{minimize}  \left\| x'-x \right\|_2^2 + c \cdot (l_1(x')+...+l_n(x'))
\end{equation}

\subsubsection{EAD}
The Elastic-Net Attacks to DNNs (EAD) was introduced by \citet{chen2018ead} as a novel approach for crafting adversarial examples. According to \citeauthor{chen2018ead}, EAD generalizes the Carlini \& Wagner \begin{math} L_2 \end{math} attack \cite{carlini2017towards} and is able to craft more effective adversarial examples based on \begin{math} L_1 \end{math} distortion metrics. Given a benign input image \begin{math} \mathbf{x_0} \end{math} and correct label \begin{math} t_0 \end{math}, the loss function \begin{math} f \end{math} for crafting EAD adversarial examples for with respect to a benign labelled input image is expressed as:

\begin{equation}
\label{ead}
    \begin{split}
        \operatorname*{minimize}_\mathbf{x} \; c\cdot f(\mathbf{x},t)+\beta\|\mathbf{x}-\mathbf{x}_0\|_1 \; + \;
        \|\mathbf{x}-\mathbf{x}_0\|_2^2 \;\; \textrm{subject to}\;\; \mathbf{x} \in [0,1]^p ,
    \end{split}
\end{equation}
where \begin{math} \mathbf{x} \end{math} is defined as the adversarial example of \begin{math} \mathbf{x_0} \end{math} with target class \begin{math} t \neq t_0 \end{math}, \begin{math} c,\beta \geq 0 \end{math} are regularization parameters of loss function \begin{math} f \end{math} and and the \begin{math} L_1 \end{math} penalty respectively. The loss function \begin{math} f(\mathbf{x},t) \end{math} for targeted attacks is defined as:

\begin{equation}
\label{ead_targeted}
    \begin{aligned}
        f(\mathbf{x},t)= \operatorname*{max} \bigg\{\operatorname*{max}_{j\neq t}\left[\mathbf{Logit(x)}\right]_j - 
        \left[\mathbf{Logit(x)}\right]_t, -\kappa \bigg\},
    \end{aligned}
\end{equation}
where \begin{math} \mathbf{Logit(x)}=[[\mathbf{Logit(x)}]_1,...,[\mathbf{Logit(x)}]_K] \in \mathbb{R}^K \end{math} is the logit layer representation of adversarial example \begin{math} \mathbf{x} \end{math} in the DNN, \begin{math} K \end{math} is the number of classes and \begin{math} \kappa \geq \end{math} is a confidence parameter.

In evaluating the performance of EAD, \citeauthor{chen2018ead} performed extensive experimental analysis using MNIST, CIFAR-10 and ImageNet datasets and found that adversarial examples generated by EAD were able to successfully evade deep neural networks trained using defensive distillation \cite{papernot2016distillation} in a similar manner as the Carlini \& Wagner attacks \cite{carlini2017towards}.

\subsubsection{Robust Physical Perturbations (RP$_{2}$)}
\citet{eykholt2018robust} introduced the  RP$_2$ method for generating adversarial examples under different physical-world conditions. This method generates adversarial perturbations which can cause targeted misclassification in deep neural networks where the adversary has full knowledge of the model. The RP$_2$ algorithm was derived in two stages: (1) using the optimization method to generate adversarial perturbation $\delta$ for single-image \begin{math} x \end{math} without considering any physical conditions, (2) update the algorithm by considering different physical conditions. The single-image optimization problem without consideration for environmental conditions is given as:

\begin{equation}
\label{robust_physical}
    \operatorname*{min}\;\; H(x+\delta,x),\;\; \textrm{s.t.}\;\;f_\theta(x+\delta) = y^* 
\end{equation}
where $f_\theta(\cdot)$ is the target classifier, $H$ is a chosen distance function, and $y^*$ is the target class. Solving Eq. \ref{robust_physical} required \citeauthor{eykholt2018robust} to reformulate it in Lagrangian-relaxed form as:

\begin{equation} 
\label{rp_lagrangian_relaxed_form}
\operatorname*{argmin}_\delta \lambda\|\delta\|_p + J\bigg(f_\theta(x+\delta),y^*\bigg)
\end{equation}
where $J(\cdot,\cdot)$ is the loss function and $\lambda$ is a hyper-parameter that controls regularization of the distortion.

After accounting for environmental conditions, the spatially-constrained perturbation optimization problem is defined as:

\begin{equation}
\label{rp_spatially_constrained}
\begin{aligned}
    \operatorname*{argmin}_\delta \lambda\|M_x \cdot \delta\|_p \;+\; \textrm{NPS} \;
        +\; \mathbb{E}_{x_i \sim X^V} J\bigg(f_\theta(x_i + T_i(M_x \cdot \delta)),y^*\bigg)
\end{aligned}
\end{equation}
where $T_i(\cdot)$ is the alignment function that maps object transformation to the perturbation transformations.


\subsubsection{Show-and-Fool}
This is a novel optimization method for crafting adversarial examples in neural image captioning proposed by \citet{chen2018attacking}. The Show-and-Fool algorithm provides two different approaches for crafting adversarial examples namely \textit{(1) targeted caption method} and \textit{(2) targeted keyword method}. The process for crafting the adversarial examples is formulated as optimization problems with objective functions that adopts the hybrid CNN-RNN architecture. For a given input image $\mathbf{x}$, an adversarial example can be obtained by solving the optimization problem:

\begin{equation}
\label{show_and_fool_optimization}
\begin{aligned}
    \operatorname*{min}_\delta c \cdot \textrm{loss}(\mathbf{x}+\delta) + \|\delta\|_2^2
        \textrm{s.t.}\;\; \mathbf{x}+\delta \in [-1,1]^n
\end{aligned}
\end{equation}
where $\delta$ is the adversarial perturbation to $\mathbf{x}$, $\|\delta\|_2^2=\|(\mathbf{x}+\delta)-\mathbf{x}\|_2^2$ is an $\ell_2$ distance metric between the benign image and the adversarial image, loss($\cdot$) is an attack loss function and term $c>0$ is a regularization constant.
For a \textit{targeted caption} denoted by $S = (S_1, S_2, ..., S_t, ..., S_N)$, where $S_t$ is the index of the $t$-th word in the vocabulary $\mathcal{V}$ and $N$ is the length of caption $S$, the loss function is given as:
\begin{equation}
\label{show_and_fool_loss_function}
\begin{aligned}
    \textrm{loss}_{S,\textrm{logits}}(\mathbf{x}+\delta)= \sum_{t=2}^{N-1}\operatorname*{max}\bigg\{-\epsilon,
        \operatorname*{max}_{k\neq S_t}\{z_t^{(k)}\}-z_t^{(S_t)}\bigg\},
\end{aligned}
\end{equation}
where $\epsilon>0$ is the confidence level which accounts for the difference between $\operatorname*{max}_{k\neq S_t}\{z_t^{(k)}\}$ and $z_t^{(S_t)}$, and $z_t := [z_t^{(1)},z_t^{(2)},...,z_t^{(|\mathcal{V}|)}] \in \mathbb{R}^{|\mathcal{V}|} $ is a vector of ${logits}$ for each possible word in the vocabulary. Based on applying the loss function of Eq. \ref{show_and_fool_loss_function} to Eq. \ref{show_and_fool_optimization}, the targeted caption method for a given targeted caption $S$ is formulated as:

\begin{equation}
\label{show_and_fool_targeted_caption}
\begin{aligned}
    \operatorname*{min}_{w \in \mathbb{R}^n} c \cdot \sum_{t=2}^{N-1}\operatorname*{max}\left\{-\epsilon,
    \operatorname*{max}_{k\neq S_t}\{z_t^{(k)}\}-z_t^{(S_t)}\right\} 
        + \| \textrm{tanh}(w+y) - \textrm{tanh}(y)\|_2^2.    
\end{aligned}
\end{equation}
Similarly for \textit{targeted keyword} denoted by $\mathcal{K} := \{K_1,...,K_M\} \subset \mathcal{V}$, the loss function is given as: 

\begin{equation}
\label{show_and_fool_targeted_keyword}
\begin{aligned}
    \sum_{j=1}^{M}\operatorname*{min}_{t \in [N]}\left\{g_{t,j}(\operatorname*{max}\{-\epsilon,
    \operatorname*{max}_{k\neq {K_j}}\{z_t^{(k)}\}-z_t^{(K_t)}\})\right\}
\end{aligned}
\end{equation} 
In evaluating the effectiveness of the Show-and-Fool method, \citet{chen2018attacking} carried out extensive experimental analysis using the Microsoft COCO dataset and the inception-v3 model. They showed that adversarial examples generated using the targeted caption and keyword methods are highly effective against neural image captioning systems and are highly transferable to other models, even those with different architectures.

\section{Defenses against AI System Attacks}
\label{defenses-ai}
\begin{figure*}[h]
\centering\includegraphics[width=0.8\linewidth]{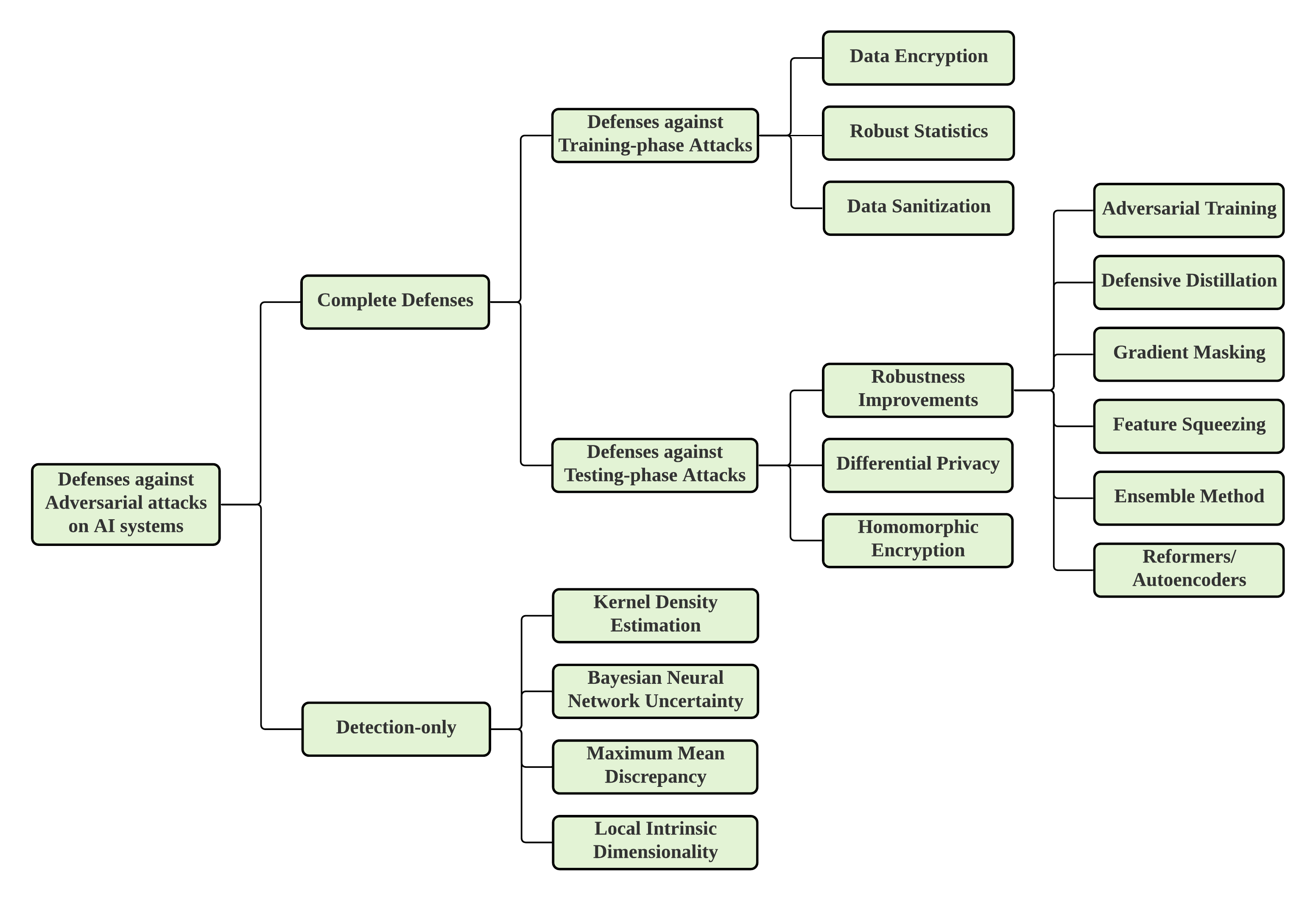}
\caption{Taxonomy of Defenses against AI system attacks}
\label{taxonomy_of_defenses}
\end{figure*}

In this section, we categorize defenses against AI system attacks based on whether they provide complete defense against adversarial examples or whether they simply detect and reject the adversarial examples. We then use this categorization, shown in Fig. \ref{taxonomy_of_defenses}, for our review of existing defense techniques. Complete defenses can be characterized by whether they apply to attacks launched against the training or testing phases of the system operation \cite{tabassi2019taxonomy}. 

\subsection{Defenses against Training Phase Attacks}
Poisoning attacks, like many other attacks on AI systems make learning inherently more difficult. Many training attack defense mechanisms are based on the assumption that poisoned samples are usually out of the expected input distribution \cite{papernot2018security}. A number of techniques have been proposed as  defenses against these training-time attacks, and some of these techniques are discussed as follows:

\subsubsection{Data Sanitization} 
Data Sanitization is a poisoning attack defense strategy that involves filtering out contaminated samples from a training dataset before using it to train a model \cite{chan2018data}. \citet{cretu2008casting} introduced a novel data sanitization technique for defending poisoning attack on out-of-the-box anomaly detection classifiers. \citeauthor{nelson2008exploiting} also presented a data sanitization technique \cite{nelson2008exploiting, nelson2009misleading} for defending against poisoning attacks known as Reject On Negative Impact (RONI). This technique was used to successfully filter out dictionary attack messages on the SpamBayes spam filter, and it accurately identified all of the attack emails without flagging any of the non-spam emails. While data sanitization techniques have been effective in defending against some data poisoning attacks developed without explicitly considering defenses, \citet{koh2018stronger} presented three new data poisoning attacks that can simultaneously evade a broad range of data sanitization defenses. The success of their work against anomaly-based data sanitization defenses simply suggests that more research effort is required in finding effective defenses against poisoning attacks.

\subsubsection{Robust Statistics} 
As opposed to data sanitization, this defense approach does not attempt to detect poisoned samples, but instead focuses on design of robust models against poisoning attacks. In their work, \citet{biggio2010multiple} investigated the use of multiple classifier systems (MCSs) for improving the robustness of pattern recognition models in adversarial settings based on the principle that more than one classifier has to be evaded in order to make the system ineffective. The results they obtained from the experimental investigations suggested that randomisation-based MCS construction techniques can be used to improve the robustness of linear classifiers in adversarial settings. Also, \citet{biggio2011bagging} experimentally demonstrated that 'bagging', an acronym for bootstrap aggregating, is an effective defense technique against poisoning attacks irrespective of the base classification algorithm. Bagging was originally proposed by \citet{breiman1996bagging} as a technique for improving classifier accuracy by generating multiple versions of a classifier and using these to get an aggregate classifier, and it has proved to be effective especially when applied to classifiers whose predictions vary significantly with little variation in training data.

\subsection{Defenses against Testing (Inference) Phase Attacks}
Defenses against testing-phase attacks include various model robustness improvements, differential privacy and homomorphic encryption. 

\subsubsection{Robustness Improvements} A machine learning model may achieve robustness by being able to detect and reject adversarial examples. While these robustness improvement techniques are defenses against testing-phase attacks, they are deployed during the training phase that precedes testing. Some of the robustness improvement techniques that have been well-researched includes Adversarial Training, Defensive Distillation, Ensemble Method, Gradient Masking, Feature Squeezing and Reformers/Autoencoders \cite{tabassi2019taxonomy}. 

\paragraph{Adversarial Training} Adversarial training seeks to improve the robustness of a machine learning model by proactively generating adversarial examples and augmenting them with training data as part of the training process. \cite{tramer2018ensemble}. These adversarial examples are perturbed examples found by optimizing the input to maximize the model's prediction error \cite{szegedy2013intriguing}. While adversarial training may be time consuming due to the number of iterative computations required to form a robust model, it has proved to be resistant to white-box attacks \cite{shafahi2019adversarial} if the perturbations computed during training closely maximizes the model's loss \cite{madry2017towards}. On the other hand, deep neural network models that are adversarially trained using fast single-step methods are found to remain vulnerable to simple black-box attacks \cite{tramer2018ensemble}.

\paragraph{Defensive Distillation} Distillation is a technique designed for transferring knowledge from an ensemble of models or large highly regularized models to smaller distilled models while preserving the prediction accuracy \cite{papernot2018security}. The technique was formally introduced by \citet{hinton2015distilling}, after being initially suggested by \citet{ba2014deep} as a method for smoothing a model's prediction accuracy by training shallow neural networks to mimic deeper neural networks for the benefit of deploying deep learning in computationally constrained devices. \citet{papernot2016distillation} also presented defensive distillation as a technique for training models that are more robust to input perturbations. Their experiments demonstrated that distilled DNN-based models are more resilient to adversarial examples. The applicability of this defense against other testing-phase attacks were also demonstrated in \cite{papernot2016limitations, goodfellow2014explaining}. Despite the benefits of defensive distillation and it's strong guarantees against adversarial examples, \citet{carlini2017towards} experimentally demonstrated how their attacks using high-confidence adversarial examples can break defensive distillation

\paragraph{Gradient Masking} 
Gradient masking is a technique that seeks to reduce the sensitivity of a DNN model to small input perturbations \cite{tabassi2019taxonomy}. In this defense strategy, the first order derivatives of a model is computed with respect to its input and these derivatives are then minimized during the learning phase \cite{papernot2018security}, resulting in a technique that conceals a model's gradient information to an adversary seeking to exploit the gradient-based information. While it is sensible to reduce a model's sensitivity to small changes in the input as a way of defending against adversarial examples, experiments performed by \citet{ papernot2016practical} revealed the flaws in the gradient masking technique by demonstrating how it can be evaded by a black-box attack. 

\paragraph{Feature Squeezing} 
This is a state-of-the-art technique for detecting adversarial examples by reducing the feature input spaces available to an adversary. It achieves this by combining many samples that correspond to different feature vectors in the original input space into a single sample.  \citet{xu2017feature} demonstrated two types of feature squeezing methods in the image space: \textit {(1) image color depth reduction} and \textit {(2) smoothening to reduce variation among pixels (median smoothening)}, as accurate and robust detection techniques against adversarial inputs. They extended this work in \cite{xu2017squeezing} where they showed that the median smoothening is the most effective squeezer in mitigating the Carlini \& Wagner attack. While feature squeezing corrupts input features that adversaries rely on for their attacks, experimental evaluations \cite{tao2018attacks} show that it could potentially cause degradation of accuracy on classifying benign inputs.

\paragraph{Ensemble Method} 
Ensemble methods are learning algorithms used to improve classification decisions of supervised learning models \cite{dietterich2000ensemble}. While many ensemble methods have been introduced in literature over the years, they have only recently been considered as a method for improving defenses of machine learning models against adversarial attacks. As a method of increasing the robustness of convolutional neural networks to adversarial attacks, \citet{abbasi2017robustness} proposed the use of an ensemble of diverse specialist classifiers to detect and reject adversarial examples while accepting benign samples based on confidence. The authors showed that this method can be used to reduce the prediction confidence for adversarial examples while preserving the confidence of clean samples to a certain extent. However, \citet{he2017adversarial} demonstrated that ensembles constructed using this method can be easily evaded by an adaptive adversary who can effectively generate adversarial examples with low distortions. \citet{strauss2017ensemble} considered ensemble methods as sole defense against adversarial attacks. Compared to existing ensemble methods, their method improves prediction confidence for clean samples while increasing robustness against adversarial examples at the cost of increased computational complexity. \citet{liu2018towards} proposed the Random Self-Ensemble (RSE), a defense algorithm that combines the concept of randomness and ensemble to improve the robustness of neural networks. Using experimental results, the authors were able to demonstrate that this method can generalize well and is an effective defense against attacks such as the Carlini \& Wagner attack \cite{carlini2017towards}.

\paragraph{Reformers/Autoencoders} A reformer is a network that takes an input \begin{math} x \end{math} and reconstructs it to \begin{math} x' \end{math} before passing it to a classifier \cite{yin2019adversarial}. An ideal reformer is not expected to change the prediction accuracy of benign examples, however it is expected to change adversarial examples such that the reformed examples are close to the benign examples. An autoencoder is an artificial neural network that comprises of an encoder and a decoder, where the encoder learns a set of hidden representation for a given input data and the decoder reconstructs the output using the hidden representation \citet{gu2014towards}. \citet{meng2017magnet} introduced MagNet, a robust defense against adversarial attacks on deep neural networks. MagNet, as a framework, incorporates a reformer and a detector. The detector, implemented using an autoencoder, calculates reconstruction error and rejects examples with high reconstruction errors. As an attack-independent defense method, experimental results show that MagNet achieved 99\% classification accuracy on adversarial examples generated by the most of the attacks considered.

\subsubsection{Differential Privacy}
\label{Differential_Privacy}
\citet{lecuyer2018connection} proposed the use of differential privacy as a means of improving the robustness of deep neural networks against adversarial attacks. In the machine learning context, differential privacy involves the introduction of randomness to training data or model outputs in order to limit the disclosure of private information of individual records included in the training data. \citet{lecuyer2019certified} introduced PixelDP, a certified defense against adversarial examples based on differential privacy. When applied to deep learning models, differential privacy helps to protect training data from inversion attacks aimed at reconstructing the training data from model parameters \cite{bae2018security}. \citeauthor{lecuyer2019certified} described PixelDP as a scalable, robust defense which applies broadly to different DNN model types. Experimental results show PixelDP to provide more accurate predictions under the euclidean norm attack compared to other certified defenses; however, it adds some computational overhead for training and testing.

\subsubsection{Homomorphic Encryption} 
\label{Homomorphic_Encryption}
Recent advances in homomorphic encryption schemes, specifically the fully homomorphic encryption, enable certain operations, such as addition and multiplication, to be carried out on encrypted data without the need to decrypt the data \cite{aslett2015review}. Homomorphic encryption has been proposed in recent literature as a means of achieving the privacy-preservation requirements for sensitive data when using predictive models managed by third-parties, such as in the case of machine learning as a service platforms. \citet{xie2014crypto} introduced crypto-nets, a neural network which makes predictions over encrypted data and results in encrypted form. Through theoretical and practical analysis, the authors demonstrated that the computational complexity involved in applying neural networks to encrypted data makes crypto-nets infeasible in specific scenarios. \citet{hesamifard2017cryptodl} introduced CryptoDL, a solution for implementing deep neural network algorithms over encrypted data. The primary components of CryptoDL are deep convolutional neural networks, trained using low degree polynomials, and homomorphic encryption. The authors designed CryptoDL to be trained using low degree polynomials as this is essential for efficient homomorphic encryption schemes and overcomes the practical limitations of other solutions that use homomorphic encryption, such as crypto-nets. Experimental results show that CryptoDL is scalable and provides efficient and accurate privacy-preserving predictions.

\subsection{Detection-only defenses} 
Detection-only defenses have recently attracted research attention due to the limitations of existing complete defenses. Works by \citet{carlini2017adversarial} and \citet{he2017adversarial} have shown that many of the complete defenses can easily be broken or evaded. The goal of detection-only defenses is to provide a distinction between clean examples and adversarial examples, and to filter out the adversarial examples for reliable classification. We now discuss some of the proposed detection-only defenses.
\subsubsection{Kernel Density Estimation (KDE) based detector}
The kernel density estimation method was proposed by \citet{feinman2017detecting} for detecting adversarial examples in deep neural networks. It uses the kernel density estimates calculated in the feature space of the last hidden layer of a neural network to detect points that lie far from the data manifold. According to \citeauthor{feinman2017detecting}, the density estimate for a point $x$ with predicted class $t$ is given as:

\begin{equation}
\label{kd_estimation}
\begin{aligned}
    \hat{K}(x,X_t)=\sum_{x_i \in X_t} k_\sigma(\phi(x),\phi(x_i))
\end{aligned}
\end{equation} 
where $\phi(x)$ is the last hidden layer activation vector for point $x$, $X_t$ is the training set of class $t$ and $\sigma$ is the tuned bandwidth. \citeauthor{feinman2017detecting} showed that this defense is effective against the FGSM, BIM, JSMA and C\&W attacks based on experiments using the MNIST, CIFAR-10 and SVHN datasets. However, \citet{carlini2017adversarial} later demonstrated how to break the KDE by using the C\&W to generate adversarial examples for MNIST with increased distortion.

\subsubsection{Bayesian Neural Network Uncertainty based detector}
The Bayesian Neural Network Uncertainty method for identifying adversarial examples was proposed by \citet{feinman2017detecting} and is used to detect points that lie in the low confidence regions of the input space. According to \citeauthor{feinman2017detecting}, this method can be used to obtain additional information about model confidence which are not normally available using methods based on distance metrics such as the KDE. The proposed bayesian uncertainty method is based on adding randomness to the neural network by using \textit{dropout}, a method introduced in \cite{srivastava2014dropout} for reducing overfitting when training deep neural networks. The uncertainty estimate of the neural network on a given instance $x^*$ and stochastic predictions $\{\hat{y_1^*},...,\hat{y_T^*}\}$ can be computed as:

\begin{equation}
\label{bayesian_uncertainty}
\begin{aligned}
    U(x^*) = \frac{1}{T}\sum_{i=1}^T \hat{y_i^*}^T\hat{y_i^*} \;\;-
        \left( \frac{1}{T}\sum_{i=1}^T \hat{y_i^*}\right)^T\left( \frac{1}{T}\sum_{i=1}^T \hat{y_i^*}\right)
\end{aligned}
\end{equation}
Based on experimental analysis using the LeNet convolutional neural network trained with dropout rate of 0.5, \citeauthor{feinman2017detecting} found the Bayesian uncertainty method to be effective in detecting adversarial examples crafted using a wide variety of attack methods. However, \citet{carlini2017adversarial} demonstrated that this method can also be vulnerable to adversarial attack.

\subsubsection{Maximum Mean Discrepancy based detector}
\citet{grosse2017statistical} proposed the use of the Maximum Mean Discrepancy (MMD), a statistical hypothesis test, to detect adversarial examples from a given input. The MMD test is based on the framework of two-sample statistical hypothesis testing used to determine whether samples $X_1$ and $X_2$ are drawn from then same distribution. If sample $X_1$ is drawn from distribution $p$ and sample $X_2$ is drawn from distribution $q$, null hypothesis $H_0$ states that $p=q$ while the alternative hypothesis $H_A$ indicates that $p \neq q$. The statistical test takes the two samples as its input and distinguishes between $H_0$ and $H_A$.

The MMD test is a kernel-based test introduced by \citet{gretton2012kernel} for use when considering data with high dimensionality. Using it as a detection technique means \citeauthor{grosse2017statistical} had to focus on the asymptotic distribution of the unbiased MMD. This algorithm involves using a subsampling method to draw samples from the data available with replacement in order to consistently estimate the distribution of the MMD under null hypothesis. 

While \citeauthor{grosse2017statistical} found that the MMD can statistically distinguish adversarial examples from clean examples, \citet{carlini2017adversarial} later demonstrated that the MMD fails to detect attacks when targeted adversarial examples crafted with the C\&W attack algorithm are used.

\subsubsection{Local Intrinsic Dimensionality based detector}
\citet{ma2018characterizing} proposed the use of Local Intrinsic Dimensionality (LID) to identify adversarial examples from a given input. In contrast to other detection-only defenses which are based on density distribution in dataset, the LID-based detector characterizes the intrinsic dimensionality of adversarial regions of deep neural networks based on local distance distribution from a reference data point to its neighbors \cite{houle2018correlation, houle2017local}. In their work, \citeauthor{ma2018characterizing}  demonstrated how well LID estimates can be used to detect adversarial examples by using the Maximum Likelihood Estimator (MLE) of LID to approximate the true distance distribution. 

Formally, given a reference data sample $x \in p$, where $p$ is the data distribution, the maximum likelihood estimator of the LID at $x$ is defined as:

\begin{equation}
\label{mle_lid}
\begin{aligned}
    \widehat{\textrm{LID}}(x) = - \left(\frac{1}{k} \sum_{i=1}^k \log \frac{r_i(x)}{r_k(x)}\right)^{-1}
\end{aligned}
\end{equation}

\citeauthor{ma2018characterizing} demonstrated how LID characteristics can facilitate the identification of adversarial examples crafted using a wide range of attack algorithms. They also showed how the characterization of adversarial regions could be used as features in an adversarial example detection process.

\section{Challenges and Future Work}
\label{challenges-future}

In previous sections, we presented a comprehensive review of security and privacy issues in AI systems covering many machine learning models. In this section, we discuss some challenges in the field and provide suggestions on future research directions.

\subsection{Transferability of adversarial examples}
Many machine learning models, including deep neural networks, are subject to the transferability of adversarial examples \cite{wiyatno2019adversarial}. The transferability phenomenon means that adversarial examples crafted to fool a particular model can easily be used to fool other models. This property poses a security challenge for many deep neural networks because adversarial examples generated from one model can be used to attack another model without knowledge of the target model's parameters \cite{yuan2019adversarial}, thereby enabling black-box attacks \cite{papernot2016transferability}. While many of the works reviewed in this paper show real evidence of transferability of adversarial examples, the fundamental reasons why adversarial examples transfer are not well understood. Based on experimental analysis in \cite{tramer2017space}, the hypothesis on the ubiquity of transferability does not always hold and there are suggestions that transferability property is not inherent in non-robust models despite existence of adversarial examples.  Therefore, research efforts focused on understanding the transferability phenomenon is essential for creating robust machine learning models.

\subsection{Evaluating robustness of defense methods}
Many of the existing defenses against adversarial attacks have already been broken or bypassed, raising concerns about the lack of thorough evaluation of the robustness of existing defense methods. In their work, \citet{carlini2017adversarial} reviewed ten existing methods for detecting adversarial examples and demonstrated how easy it is for an adversary to bypass these methods by constructing new cost functions. Their work shows that existing defenses are not robust against adaptive adversaries while also lacking thorough security evaluations. In Section \ref{Differential_Privacy} we discussed Differential Privacy as a method for improving the robustness of deep learning models against inversion attacks aimed at reconstructing training data from model parameters. Differential Privacy adds randomness to the training data as a means of limiting the disclosure of private information included within the training dataset. While the efficacy of differential privacy has been demonstrated in a number of experiments, there are no practical evaluations or metrics that can be used to measure if the differential privacy bounds are strong enough \cite{bae2018security}. Also in Section \ref{Homomorphic_Encryption}, we discussed homomorphic encryption as an encryption scheme that can be used to preserve privacy of sensitive data used in many predictive models. While recent implementations of homomorphic encryption such as CryptoDL \cite{hesamifard2017cryptodl} have demonstrated high prediction rates, the prediction accuracy is not at par with many state-of-the-art deep models and is found to be incompatible with deeper models.  A future research direction should, therefore, seek to investigate the properties that defense methods need to have in order to guarantee robustness against adaptive adversaries.

\subsection{Difficulty in controlling the magnitude of adversarial perturbations}
The different methods for generating adversarial examples impose small imperceptible input perturbations in order to change a neural network's prediction. However, determining the exact magnitude of perturbations required to fool a neural network is difficult, because input perturbations that are too small cannot generate adversarial examples, and perturbations that are too large are not imperceptible \cite{zhang2019adversarial}. Therefore, being able to control the magnitude of input perturbations poses an ongoing challenge.

\subsection{Lack of research focus on attacks beyond classification tasks}
Convolutional neural networks have been hugely successful in computer vision applications. As a result, a majority of the existing methods for generating adversarial examples apply to computer vision applications such as image recognition and object detection which are part of the machine learning classification tasks \cite{papernot2016limitations, kurakin2016adversarial, goodfellow2014explaining, sarkar2017upset, moosavi2016deepfool, moosavi2017universal}. While some attention has been given to adversarial attacks on other machine learning tasks, these are only a handful in comparison to the attention given to classification tasks. It has been established in many literature that all machine learning models, including deep neural networks, are vulnerable to adversarial attacks. Therefore, more research effort on the adversarial threats facing other machine learning task categories, such as reinforcement learning, is required.

\subsection{Evolving threat of unknown unknowns}
 Unknown unknowns pose a significant threat to machine learning systems deployed in adversarial environments, similar to how they are a real threat in many cybersecurity problems such as malware and intrusion detection \citet{biggio2018wild}. Unlike, known unknowns which are used to model attacks in adversarial machine learning, unknown unknowns are often unpredictable and can cause machine learning models to misclassify with high-confidence due to inputs that are significantly different from the known training data. For machine learning systems to have the capability to detect unknown unknowns using robust techniques for anomaly detection, new research paths would need to be explored in this area.

\subsection{Randomization of classifier's decision boundary}
Introducing some randomization in the placement of the decision boundary for classifiers has been proposed as a method for improving the classifier's security against evasion attacks \citet{barreno2006}. Surprisingly, there has only been a few attempts in literature at investigating this kind of randomization as a viable defense method. While randomization does indeed increase the amount of work that an adversary would need to do in order to move the decision boundary past a targeted point, \citeauthor{barreno2006} acknowledged that a major challenge with using randomization is that it can also increase the classifiers initial error rate, thereby degrading its performance on clean data. The problem of finding an appropriate amount of randomization that can be introduced into a model to achieve robustness against adversarial attacks remains an open problem. 

\section{Conclusion}
\label{S:8}

The field of adversarial machine learning has received significant research attention in recent years. In particular, many research works have explored adversarial attacks on machine learning models in the context of computer vision and image recognition, natural language processing and cybersecurity. Several defense methods have also been proposed with research efforts focused on evaluating the effectiveness of these defense methods against the continually evolving adversarial attacks.
 
In this work, we started with a review of recent surveys papers focusing on the issues of security and privacy in AI. We find that many of these existing surveys did not cover attacks and defenses across all machine learning task categories, with most only focusing on deep neural networks in the context of computer vision, natural language processing and cybersecurity. As a basis for describing adversarial attacks on machine learning models, we start by providing a theoretical background of machine learning task categories and make a clear distinction between shallow learning methods and the more recent deep learning methods. We then present a new framework for the holistic review of adversarial attacks on AI systems by first describing an adversary's goals, knowledge and capabilities, followed by a comprehensive analysis of adversarial attacks and defense methods covering many machine learning models. 

\bibliographystyle{ACM-Reference-Format}
\bibliography{bibliography}










\end{document}